\documentclass[12pt]{article}
\usepackage{epsf}
\usepackage{graphicx}
\usepackage{amssymb}

\def\di{\displaystyle}
\def\&{&\di}
\def\bg{\begin{eqnarray}\begin{array}{rcl}\displaystyle}
\def\bm#1{\begin{eqnarray}\begin{array}{#1}\displaystyle} 
\def\eg{\end{array} &\di    &\di   \end{eqnarray}}

\def\bgo{\begin{eqnarray*}\begin{array}{rcl}\displaystyle}
\def\ego{\end{array} &\di    &\di \nonumber  \end{eqnarray*}}

\def\btensor#1#2{\renew\left#1\begin{array}{#2}\di}
\def\etensor#1{\end{array}\right#1}

\def\d{{\mbox d}}

\def\Tr{\mbox{Tr}}

\def\CC{{\cal C}}
\def\CD{{\cal D}}

\def\CF{{\cal F}}

\def\CP{{\cal P}}
\def\CS{{\cal S}}

\def\CV{{\cal V}}

\date{\today}

\def\rene{\renewcommand{\arraystretch}{1.8}}
\def\renew{\renewcommand{\arraystretch}{1}}

\rene

 \newcommand{\mysection}[1]{\section{#1}%\setcounter{figure}{0}
                    \setcounter{table}{0}\setcounter{equation}{0}}

\def\D{{\mbox{D}}}
\def\ad{{\mbox{ad}}}

\begin{document}

\begin{titlepage}
\begin{flushright}
\normalsize UNITU--THEP--16/2000
\end{flushright}

\vspace{2cm}

\par
\vskip .5 truecm
\large \centerline{\bf Abelian and center gauges in continuum 
Yang-Mills-Theory\footnote{supported by DFG under grant-No. DFG-Re
856/4-1 and DFG-EN 415/1-2}} 
\par
\vskip 1 truecm
\normalsize
\begin{center}
{\bf H.~Reinhardt}\footnote{e--mail: \tt  reinhardt@uni-tuebingen.de} and 
{\bf T.~Tok}\footnote{e--mail: \tt tok@alpha6.tphys.physik.uni-tuebingen.de}
\\
\vskip 1 truecm
\it{Institut f\"ur Theoretische Physik, Universit\"at T\"ubingen\\
Auf der Morgenstelle 14, D-72076 T\"ubingen, Germany}
\\
\today

\end{center}
 \par
\vskip 2 truecm\normalsize
\begin{abstract} 
Abelian and center gauges are considered in continuum Yang-Mills theory in 
order
to detect the magnetic monopole and center vortex content of gauge field
configurations. Specifically we examine the Laplacian Abelian and center
gauges, which are free of Gribov copies, as well as the center gauge analog of
the (Abelian) Polyakov gauge. In particular, we study meron, instanton and
instanton-anti-instanton field configurations in these gauges and
determine their monopole and vortex content. 
While a single instanton does not give rise to a center vortex,
we find center vortices for merons. Furthermore we provide evidence, that
merons can be interpreted as intersection points of center vortices. For the
instanton-anti-instanton pair, we find a center vortex enclosing their
centers, which carries two monopole loops.
\end{abstract}



\end{titlepage}
\baselineskip=20pt

\mysection{Introduction}

At present there are two popular confinement mechanisms: the dual Meissner
effect \cite{Parisi,Mandelstam,tHooft-76-1}, 
which is based on a condensation of magnetic monopoles 
in the QCD vacuum and the vortex condensation picture
\cite{tHooft-78,Mack-79}. 
Both pictures were proposed long time ago,
but only in recent years mounting evidence for the realization of these 
pictures
has been accumulated in lattice calculations. The two pictures of confinement
show up in specific partial gauge fixings.

Magnetic monopoles arise as gauge artifacts in the so-called Abelian gauges
proposed by 't~Hooft \cite{tHooft-81}, where the Cartan subgroup $H$ of the  
gauge group $G$ is
left untouched, fixing only the coset $G/H$. To be more precise the magnetic
monopoles explicitly show up only after the so-called Abelian projection, which
consists in throwing away the ``charged'' part of the gauge 
field after implementing
the Abelian gauge. Magnetic monopoles appear at those isolated points in space,
where the residual gauge freedom is larger than the Abelian subgroup. 

Since the magnetic monopoles arise as gauge artifacts, their occurrence and
properties depend on the specific form of the Abelian gauge used. For example,
monopole dominance in the string tension 
\cite{Suzuki-90,Hioki-91,Bali-98} 
is found in maximally Abelian gauge, but not in Polyakov gauge 
\cite{Bernstein-96} (in Polyakov gauge there is, however, an exact
Abelian dominance in the temporal string tension). However, in all forms of the
Abelian gauges considered monopole condensation occurs in the confinement phase
and is absent in the de-confinement phase \cite{DiGiacomo-00}. 

The vortex picture of confinement, which received rather little attention after
some early efforts following its inception has recently received strong support
from lattice calculations performed in the so-called maximum center
gauge \cite{DelDebbio-97,DelDebbio-98}, where
one fixes only the coset $G/Z$ but leaves the center $Z$ of the gauge group $G$
unfixed\footnote{The continuum analog of the maximum center gauge has 
been derived in ref. \cite{Engelhardt-00-1}.}.
Subsequent center projection, which consists in replacing each link by
its closest center element allows the identification of the center vortex
content of the gauge fields. Lattice calculations show, that the vortex 
content detected after center projection produces virtually the full 
string tension, while the string tension disappears, if the center 
vortices are removed from the lattice ensemble 
\cite{DelDebbio-97,Forcrand-99-1}. 
This fact has been referred to as center dominance.
Center dominance persists at finite temperature 
\cite{Langfeld-99,Engelhardt-00-2} for both the $q - \bar
q$ potential (Polyakov loop correlator) as well as for the spatial
string tension. The vortices have also been shown to condense in the 
confinement phase \cite{Kovacs-00}. Furthermore in  the gauge field 
ensemble devoid of center vortices chiral symmetry breaking disappears 
and all field configurations belong to the topologically trivial sector
\cite{Forcrand-99-1}.

Unfortunately, both gauge fixing procedures, the maximally Abelian gauge 
and the maximum center gauge, suffer from the Gribov problem
\cite{Gribov-78}, both 
on the lattice as well as in the continuum \cite{Bruckmann-00-1}. 

To circumvent the Gribov problem, the Laplacian gauge \cite{Vink-92},
the Laplacian Abelian gauge \cite{vdSijs-96}  
and the Laplacian center gauge 
\cite{Alexandrou-99-1,Alexandrou-99-2}  
have been introduced. 
In the latter two gauges, which are free of Gribov copies, one uses 
eigenvectors of the covariant Laplace 
operator in the adjoint representation to fix the gauge. 
These eigenvectors  
transform homogeneously under gauge rotations. 
In the Laplacian Abelian gauge fixing one exploits the gauge freedom
to rotate the lowest eigenvector of the covariant Laplacian into the Cartan
subalgebra. In the Laplacian center gauge, one uses the residual Abelian 
gauge freedom, which remains after Laplacian Abelian gauge fixing to 
rotate the next to lowest eigenvector into the plane spanned by the first 
and third axes in color space (for gauge group $SU(2)$). 
The Laplacian center gauge fixing has 
the advantage, that the magnetic monopoles lie on the vortices by construction.

In this paper we consider various types of Abelian and center gauges in
continuum Yang-Mills theory and study in these gauges field
configurations which are considered to be relevant in the infrared
sector of QCD like center vortices, instantons and merons. Center
vortices can give an appealing explanation of confinement (see 
e.g.~ref.~\cite{Engelhardt-00-2}). 
It is the general consense that instantons have
little to do with confinement but offer an explanation of spontaneous
breaking of chiral symmetry \cite{Callan-78,Callan-79}. Merons can be
considered as ``half of an instanton with zero radius'' and we will
provide evidence that they can be regarded as intersection points of
center vortices.

The advantage of the Abelian and center gauges is that they provide a
convenient tool to detect the monopole and vortex content of a field
configuration.

Previously the monopole content of instantons has been considered in the
Polyakov gauge and maximally Abelian gauge 
\cite{Chernodub,Hart,Brower,Bornyakov}. 
In maximally Abelian gauge
a monopole trajectory was found to pass through the center of the
instanton in ref. \cite{Chernodub}, while an infinitesimal monopole loop
around the center of the instanton was found in \cite{Brower}. 
These results are consistent with the findings of \cite{Bruckmann-00-2}
where an instanton on an $S^4$-space-time manifold has been considered
and a monopole loop degenerate to a point was found in Laplacian Abelian
gauge. Only for a special
choice of the instanton scale one can find a monopole loop of finite size
\cite{Bruckmann-00-2}. In
Polyakov gauge \cite{Weiss-81} a static monopole trajectory passes 
through the center of the instanton \cite{Suganuma-95}. In this gauge the
Pontryagin index can be entirely expressed in terms of magnetic monopole
charges \cite{Reinhardt-97-2,Jahn-98,Ford-98-1,Quandt-98}. 

The vortex content of instanton field configurations has been less
understood. The first investigations in this direction have been
reported in ref. \cite{Alexandrou-99-1,Alexandrou-99-2} 
where a cooled two instanton configuration and a cooled caloron
configuration have been considered in the Laplacian center gauge on the
lattice. 
In the former case a vortex sheet was found connecting the positions of
the two instantons. In the case of the caloron which can be interpreted
as a monopole-anti-monopole pair the vortex sheet runs through
the positions of monopole and anti-monopole, which is expected since in
the Laplacian center gauge by construction the monopoles are sitting on
the vortex sheets. One should, however, keep in mind that the lattice
result cannot be straightforwardly transferred to the continuum. Due to
the periodic boundary conditions a localized configuration on the
lattice corresponds to an array of such configurations in the continuum.
In addition the detection of topological charge on the lattice is
problematic on its own. 

In this paper we study various field configurations like vortices, 
instantons and
merons in the continuum analog of the Laplacian center gauge.
The organization of the paper is as follows:
In section \ref{general-gauge} we define Abelian and center gauges by
means of one and two, respectively, color fields transforming
homogeneously under gauge transformations. In section
\ref{examples} we discuss the Laplacian Abelian and center gauges in
continuum Yang-Mills-theory and provide examples for field
configurations giving rise to magnetic monopoles and center vortices
after Laplacian Abelian and center gauge fixing, respectively. In
section \ref{merons} merons, instantons and instanton-anti-instanton
pairs are considered in the Laplacian Abelian and center gauges. We also
provide evidence that merons can be interpreted as vortex intersection
points. Some concluding remarks are given in section \ref{conclusion}.

\mysection{Abelian and center gauges}
\label{general-gauge}

In the following we consider Abelian and center gauges from a general
point of view.

Since the center $Z$ of a group $G$ belongs to its Cartan subgroup $H$
(center) gauge fixing can be formed in two steps: First one fixes the
coset $G/H$ leaving the Cartan subgroup $H$ unfixed, which is referred
to as Abelian gauge fixing. Secondly one fixes the coset $H/Z $ leaving
the center unfixed, which is referred to as center gauge fixing.
In this paper we will mainly concentrate on $G=SU(2)$ and $H=U(1)$. 
For a recent generalization of the Laplacian center gauge fixing 
to the gauge group $SU(N)$ see ref.~\cite{deForcrand-00}.

\subsection{Abelian gauge}
\label{Abelian-gauge}

In this section we will shortly discuss Abelian gauges. 
It is a gauge fixing 
procedure which fixes the gauge group $G$ 
up to its Cartan subgroup 
$H$. 
For the Abelian gauge fixing one considers a Lie algebra 
valued field $ \psi_1 $ in 
the adjoint representation transforming homogeneously under gauge
transformations. In the following we will refer to such a field as
``Higgs field''.
This field can be given as the solution to some covariant field 
equation or as 
the extremum of some gauge independent functional. 
Examples will be given in 
section \ref{examples}.

Now we fix the gauge by demanding that $ \psi_1 (x) $ points in the 
$ 3 $-direction of color space for every $ x \in M $, $M$ being the 
space-time manifold, i.e.~we are searching for a gauge transformation 
$ V $ such that
\begin{equation}
\label{cond1}
V(x)^{-1} \psi_1(x) V(x) = h(x) \sigma_3 \, , \quad h(x) \geq 0 \, .
\end{equation} 
For $ \psi_1(x) \neq 0 $ the transformation matrix $ V(x) $ is defined
up to a residual $ U(1) $ gauge transformation 
$ V(x) \rightarrow V(x) \exp ( \alpha(x) \sigma_3 / (2 i) )$.
However, if $ \psi_1(x) = 0 $, then $ V(x) $ is arbitrary and the residual 
gauge 
freedom is enlarged to the full gauge group $SU(2)$. 
At such points the function $V(x) $ will in general 
become singular. This leads us to the definition of the 
Abelian defect manifold
\begin{equation}
\CD_A := \left\{ x \in M ; \psi_1 ( x ) = 0 \right\} \, .
\end{equation}
Any connected one-dimensional subset of $\CD_A$ can be 
identified with a magnetic monopole 
loop with respect to the residual $U(1)$ gauge freedom. 
This is because for $G=SU(2)$ the condition 
$\psi_1(x) = \psi_1^a \sigma_a = 0 $ implies 3 
equations $\psi_1^{a=1,2,3} (x) = 0$ which generically define 
a one-dimensional manifold in $D=4$, the monopole trajectory.
On $\CD_A^c := 
M \setminus \CD_A$ we define the Abelian magnetic gauge potential 
\cite{tHooft-81,Kronfeld-87}
\begin{equation}
\label{A-mag}
A_{{\mathrm mag}}:= \Tr ( V^{-1} \d V T_3 ) \, , \quad T_3 = \sigma_3/(2i)  
\end{equation} 
and find for its field strength
\begin{equation}
\label{F-mag}
F_{{\mathrm mag}}:=\d A_{{\mathrm mag}}=
-\Tr(V^{-1} \d V \wedge V^{-1} \d V T_3 ) \, .
\end{equation}
The field strength $ F_{{\mathrm mag}} $ does not depend on the special 
choice of
$V$, which is defined only up to a $U(1)$ gauge transformation. 
Surrounding a monopole by a 
surface\footnote{In $D=4$ the monopoles form one-dimensional lines. 
To define the surface $\CS$ we split the space-time locally into the 
monopole trajectory and three-dimensional slices transversal to the
monopole trajectory. 
Now we choose $\CS$ as a two-sphere surrounding the monopole
in a fixed three-dimensional slice.}
$\CS$ and integrating $F_{{\mathrm mag}}$ over $\CS$ results in the magnetic 
charge $q$ inside $\CS$:
\begin{equation}
q = \frac{1}{2 \pi} \int_\CS F_{{\mathrm mag}} \, .
\end{equation}
Geometrically the monopole charge $q$ is nothing but the winding number 
of the  map $\hat \psi_1 :  \CS \rightarrow SU(2) / U(1) \equiv S^2 $
\cite{Arafune-75}. 
The image $ S^2$ is given by the unit color vectors in the Lie 
algebra $su(2)$, obtained by normalizing the Higgs field:
$\hat \psi_1(x) = \psi_1(x)/|\psi_1(x)|$.
If $q\neq 0 $ it follows that $V$ cannot be chosen smoothly on $\CS$. 
$V$ has to be singular on a Dirac string emanating from the monopole inside 
$\CS $, and the Dirac string punctures $\CS$. 
The explicit location of the Dirac 
strings  is arbitrary --- the only requirement is that the total charge 
of a connected monopole-Dirac-string-network has to vanish. Everywhere 
outside the Dirac strings and the monopoles we can choose $ V$ smoothly.

Near its zeros the Higgs field of a charge one monopole looks in an 
appropriate choice of coordinates like a hedgehog:
\begin{equation}
\label{hedgehog}
\psi_1 (x) = \sum_{i=1}^3 x^i \sigma_i \, .
\end{equation}
as follows by a Taylor expansion of $\psi_1(x)$ around its zeros. 
The hedgehog is diagonalized by the matrix 
$V=\exp(\vartheta \sigma_\varphi/(2i))$
for which the Abelian part of the induced gauge field $ V^{-1} \d V $ 
develops a Dirac monopole:
\begin{eqnarray}
\label{monopole-gauge-trf}
V^{-1} \d V &=& \frac{\sigma_\varphi}{2i} \d \vartheta 
- \left( \sin \vartheta \cos \varphi \frac{\sigma_1}{2i} 
+ \sin \vartheta \sin \varphi \frac{\sigma_2}{2i} \right) \d \varphi
- (1-\cos\vartheta) \frac{\sigma_3}{2i} \d \varphi \, ,
\\
\label{monopole-gauge-pot}
A_{{\mathrm mag}} &=& \frac{1}{2} (1-\cos\vartheta) \d \varphi \, ,
\end{eqnarray}
where we used spherical coordinates in ${\mathbb R}^3$.
Its magnetic charge is $q=1$ which is equal to the winding number of the 
normalized Higgs field $\hat \psi = \hat x $ around $x=0$.

From a geometrical point of view the set of all matrices $V(x)$, 
$x \in \CD^c_A $ rotating $\psi_1(x)$ into the $\sigma_3$ 
direction defines a principal bundle $ P_A $ over $ \CD^c_A $. 
A smooth gauge transformation $V$ is a global section in 
$P_A$. On the other hand, the existence of magnetic monopoles 
tells us that the bundle 
$P_A$ is not trivial and we have to introduce Dirac strings 
on which a global section in the bundle is singular.

\subsection{Center gauge}
\label{Center-gauge}

In this section we will discuss center gauges fixing the gauge 
group up to its center. These gauges are extensions of Abelian 
gauges discussed in the previous section. To this end we consider 
a second Higgs field 
$\psi_2(x) $ in the adjoint representation. Again it should be given 
as the solution to some covariant field equation or as the 
extremum of some gauge independent functional (examples are given in 
section \ref{examples}).

After having fixed the gauge group up to its Cartan subgroup, we use the 
remaining Abelian gauge freedom $g(x) \in H $ to rotate $ \psi_2(x) $ 
into the plane spanned by 
$ \sigma_3 $ and $\sigma_1$ in color space:
\begin{eqnarray}
\nonumber
%\psi_1^{V g} (x) = V^{-1}(x) \psi_1(x) V(x) &=& h(x) \sigma_3 , 
%h(x) \geq 0 ,\\
\psi_2^{V g} (x) = g ^{-1} V^{-1}(x) \psi_2(x) V(x) g 
    &=& l_3(x) \sigma_3 + l_1(x) \sigma_1 , 
\\
\label{cond2}
    && l_1, l_3 \in {\mathbb R} \, , 
    \quad l_1(x) \geq 0 \, , 
    \quad g \in H \, .
\end{eqnarray}
Alternatively one can express this condition by saying that the part of
the color vector $\psi_2 (x)^V$ perpendicular to the $3$-direction, 
${\psi_2}_\perp$, is rotated into the $1$-direction.
As long as $\psi_1(x) $ and $\psi_2(x)$ are linearly independent the 
conditions (\ref{cond1},\ref{cond2}) fix $V(x)g(x)$ up to a factor 
$ \pm 1 $, i.e.~up to the center of 
$SU(2)$. But, if $ \psi_1(x) $ and $\psi_2(x) $ are linearly dependent, then 
the residual gauge freedom is enlarged. At such points the gauge 
transformation $ V g$ will in general be singular. This brings us to the 
definition of the center defect manifold
\begin{equation}
\CD_C := \left\{x \in M ; \psi_1(x) \, \mbox{and} \, \psi_2(x) \, 
\mbox{are linearly dependent} \right\} \, .
\end{equation}
A connected two-dimensional subset of $\CD_C$ will be called vortex. 
There are two conditions
to be fulfilled for the linear dependence of the color vectors 
$ \psi_1(x) $ and $\psi_2(x)$.
Therefore the vortices have co-dimension 2, i.e.~they are one-dimensional 
lines in $D=3$ and two-dimensional sheets in $D=4$.
By definition 
$\CD_A \subset \CD_C $ (if $\psi_1(x) = 0$, then $\psi_1(x) $ and $\psi_2(x)$ 
are obviously linearly dependent). Therefore the monopoles 
identified in the Abelian gauge lie on top of the 
vortices identified in the center gauge.

We should emphasize that the vortices identified in the center
gauge in the way described above correspond to the ideal center vortices
obtained in the maximal center gauge after center projection. 
Like in the latter case the
vortices identified in the above described center gauge carry flux 
which is not identical to the
flux carried by the original gauge field. 

\begin{figure}
\begin{minipage}{7cm}
\centerline{\epsfxsize=7 cm\epsffile{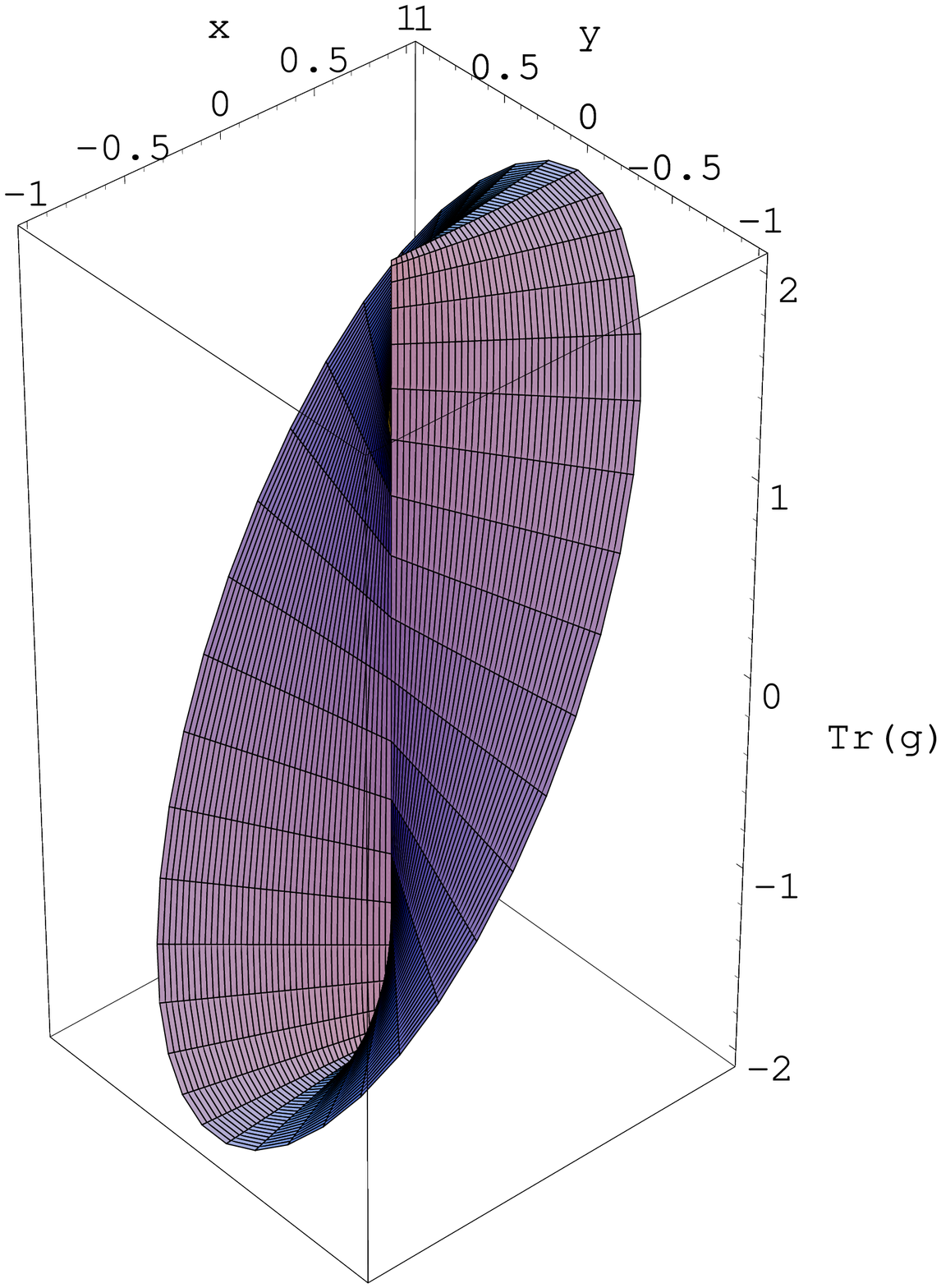}}
\caption{\label{dirac-string}\textsl{Plot of the trace of the gauge
transformation $g$, 
see equ. (\ref{vortex-gauge-trf}), for $n=2$ and positive sign. 
There is a Dirac string at $x=y=0$.
The gauge transformation can be chosen smoothly everywhere except on the
Dirac string.}}
\end{minipage}
\hspace{1cm}
\begin{minipage}{7cm}
\centerline{\epsfxsize=7 cm\epsffile{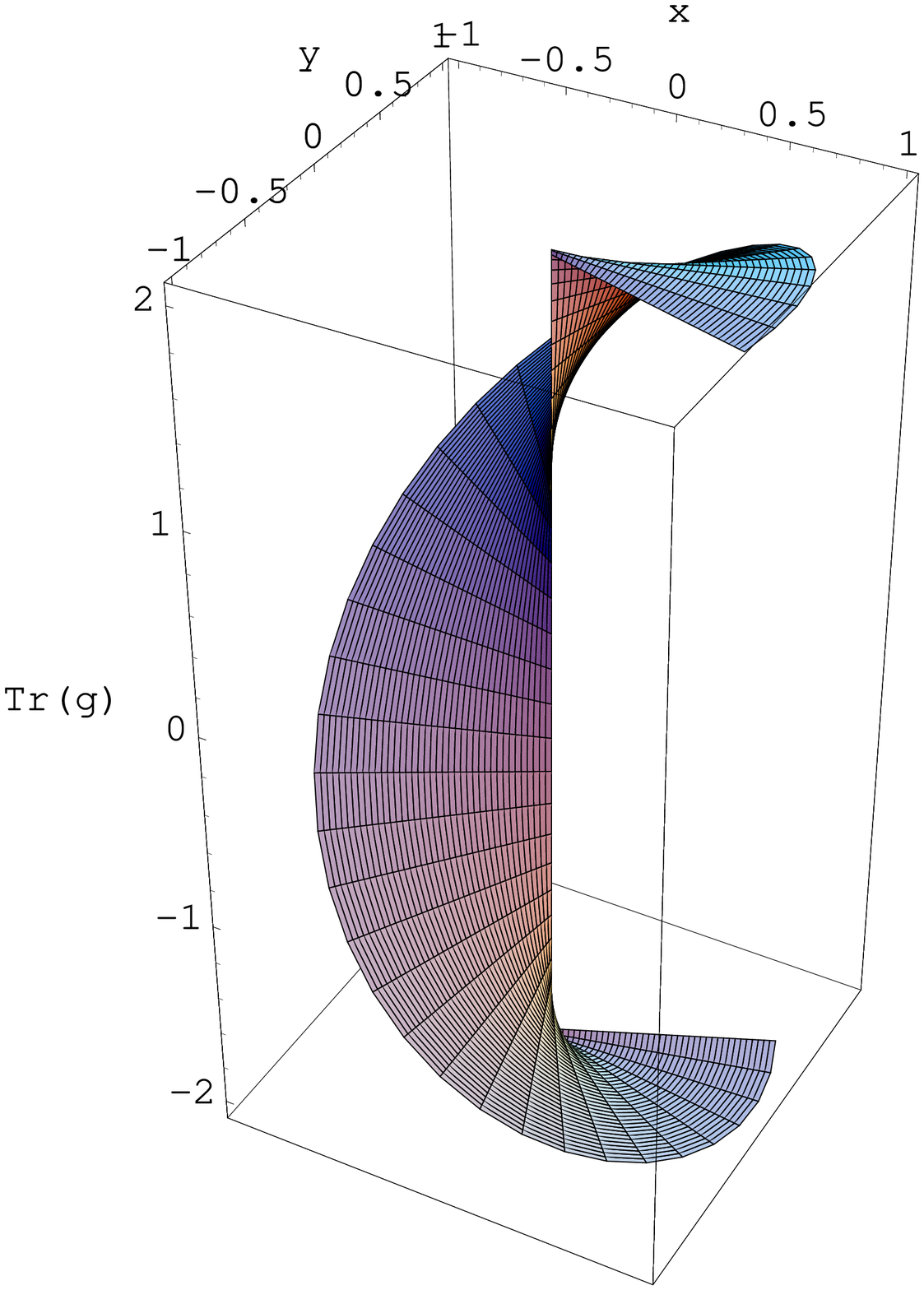}}
\caption{\label{center-vortex}\textsl{Plot of the trace of the 
gauge transformation $g$, see equ. (\ref{vortex-gauge-trf}), 
with $n=1$ and positive sign. 
There is a center vortex at $x=y=0$. The gauge 
transformation $g$ has to jump by $-1$ 
on a cut emanating from 
the vortex. At the vortex the gauge transformation $g$ is obviously 
singular.}}
\end{minipage}
\end{figure}

As an illustrative example consider the following Higgs fields in $D=3$
\begin{equation}
\label{psi1-psi2}
\psi_1 = \sigma_3 \, , \, \psi_2 ( \rho, \varphi , z) = 
\sigma_3 + \rho ( \cos (n \varphi) \sigma_1 + 
\sin (n \varphi) \sigma_2 ) \, , \quad 
n \in {\mathbb Z} \, , 
\end{equation}
where we used polar coordinates.
The gauge transformation which brings the field into the center 
gauge ($g^{-1} \psi_2 g = \sigma_3 + \rho \sigma_1$) reads
\begin{equation}
\label{vortex-gauge-trf}
g(\rho, \varphi, z) = \pm \exp ( n \varphi T_3 ) \, , \quad T_3 =
\sigma_3/(2i) \, .
\end{equation}
For even $n$ this gauge transformation is smooth everywhere except on
the $z$-axis ($\rho=0$), where $\varphi$ is ill-defined, see figure
\ref{dirac-string}.
However, for odd $n$ the gauge transformation $g$ 
becomes double-valued, i.e.~we have to introduce a cut 
(emanating from the $z$-axis) on which $g$ jumps by $-1$%
, see figure \ref{center-vortex}%
. 
The line singularity on the $z$-axis represents a Dirac string for even 
$n$ and a center vortex for odd $n$.

On the $z$-axis $\psi_2 = \sigma_3$, see (\ref{psi1-psi2}),
and hence $\psi_2$ is here parallel
to $\psi_1$. According to the previously given definition of center
vortices in the center gauge the $z$-axis hosts a center vortex.
Indeed for odd $n$ the line singularity in $g$ on the $z$-axis induces a
center vortex in the center gauge transformed potential 
$A^g = g^{-1} A g + g^{-1} \d g$.  However for even $n$ the line 
singularity in the gauge
transformation $g$ on the $z$-axis gives rise to a Dirac string in 
$g^{-1} \d g$. To see this we calculate the magnetic flux carried by 
$A_{{\mathrm mag}}$
(\ref{A-mag}) through an infinitesimal loop $\CC$ encircling the $z$-axis.
Using (\ref{vortex-gauge-trf}) we find
\begin{eqnarray}
\nonumber
\Phi &=& \frac{1}{2 \pi} \oint_\CC A_{{\mathrm mag}} =
\frac{1}{2 \pi} \oint_\CC \Tr ( (Vg)^{-1}  \d (V g ) 
T_3) 
\\ 
\nonumber
&=& 
\frac{1}{2 \pi} \oint_\CC \Tr(g^{-1} V^{-1} ( \d V g + V \d g ) 
T_3)
= \frac{1}{2 \pi} \oint_\CC \Tr(g^{-1} \d g  T_3) 
\\
\nonumber
&=&
\frac{1}{2 \pi} \oint_\CC \Tr( ( n T_3 \d \varphi) T_3) =
- \frac{1}{2 \pi} \oint_\CC n / 2 \d \varphi \\
&=& - n / 2 \, .
\end{eqnarray}
The term with $V^{-1} \d V$ in the second line vanishes in the limit 
of an
infinitesimal loop $\CC$, because $V$ can be chosen smoothly on the
vortex (away from monopoles). 
For $n$ even (odd) we find integer (half-integer) flux carried by a
Dirac sheet (center vortex sheet). 
Furthermore, the Wilson loop 
$ \CP \exp ( - \oint_\CC (Vg)^{-1} \d (Vg) ) $ becomes $-1$ (in general
a non-trivial center element) for center 
vortices and $+1$ for Dirac sheets. Thus a Dirac sheet is indeed not
seen, i.e.~unobservable by the Wilson loop, as it should since it is a
gauge artifact.

From the above considerations it is clear that we can interpret 
Dirac sheets as two center vortex sheets
on top of each other. Then it is obvious that at a monopole there 
are at least two center vortices coming in. 

The important lessen from the study of the above example is that the
procedure of identifying center vortices in the center gauge
catches not just center vortices but {\em also} Dirac strings. 

However, the generic case is the occurrence of 
center vortices. To show this let us consider a sheet singularity $\CV
\subset \CD_C $ and a point 
$ x_0 $ on it, i.e.~$ \psi_1(x_0) \sim \psi_2(x_0) $. Further we demand
$ \psi_1(x_0) \neq 0 $ and $ \psi_2(x_0) \neq 0 $. We choose 
a 2-dimensional face $\CF$ through $x_0$ which is vertical to $\CV$, 
i.e.~$\CF$ traverses $\CV$ only at $x_0$. We make a 
gauge transformation $V$ which rotates $\psi_1$ into the $\sigma_3$ 
direction. This gauge transformation can be chosen smoothly in a
neighborhood of $x_0$, because $\psi_1(x_0) \neq 0$. 
We consider 
the gauge transformed normalized field $ \hat \psi_2^V $ on $\CF$.
The field $ \hat \psi_2^V$ takes values in the unit sphere in 
$su(2) \equiv {\mathbb R}^3$, i.e.~on a two-sphere $S^2$. 
Hence it makes sense to calculate the functional 
determinant of the map $ \hat \psi_2 : \CF \to S^2 $ at the point $x_0$.
Generically this determinant is nonzero\footnote{Demanding that the 
determinant vanishes is one further condition on $ \psi_1 (x_0)$
 and $\psi_2(x_0)$. Therefore the set of such points would have dimension 
1 in $D=4$. But on the other hand we can choose $x_0 $ arbitrarily on 
$\CV$ which has dimension 2. Hence the generic case is a non-vanishing 
functional determinant.}. For non-vanishing functional determinant, using 
Taylor expansion, we can introduce 
coordinates $x^1, x^2$ on $\CF$ such that in an infinitesimal neighborhood 
of $x_0$ the field $\hat \psi_2^V(x)$ looks like 
$$
\hat \psi_2^V(x) = \pm \sigma_3 + (x-x_0)^1 \sigma_1 +  
(x-x_0)^2 \sigma_2 \, .
$$
Changing into polar coordinates on $\CF$ the fields $\psi_1^V , \psi_2^V$ 
have the form (\ref{psi1-psi2}) with $n=1$.
Completing the center gauge fixing results obviously in a center vortex at 
$x_0$.

We can again visualize the gauge fixing geometrically in a bundle picture. 
Appending to 
each $x \in \CD_C^c = M \setminus \CD_C $ the two matrices $ \{V(x)g(x), 
-V(x)g(x)\}$, which  fulfill (\ref{cond1},\ref{cond2}) we get a principal 
bundle $ P_C$ with structure group ${\mathbb Z}_2$.
The bundle $P_C$ is a twofold covering manifold of $\CD_C^c$. 
In analogy to complex function theory of a double-valued function 
(e.g.~$\sqrt{z}$) we may look at $P_C$ as the Riemann surface of 
a double-valued 
function. The vortices can be identified as  branching points. 
This gives us the opportunity to classify the line (surface)
singularities in $D=3$ ($D=4)$. We consider a closed 
loop surrounding the singularity and lift this loop into the covering manifold 
$P_C$. There are two classes of lifted loops - they can be closed (remain on 
the same sheet of the Riemann surface) or open (change the sheet 
of the Riemann surface). 
If the lifted loop is closed, the singularity represents a Dirac string. 
If the lifted loop is open (i.e.~it jumps by $-1 $ at the endpoint), 
the singularity represents a center vortex, compare also with figures
\ref{center-vortex} and \ref{dirac-string}.

We are interested in a globally well defined gauge transformation $ V $ 
on $\CD_C^c$. 
But for this we have to introduce cuts emanating from the center vortices. 
At these cuts the gauge transformation $ V $ jumps by 
the center element $ -1 $. If one would work with gauge group $ SO(3) $ from 
the very beginning these cuts would be invisible, because 
$ SO(3) \equiv SU(2) / {\mathbb Z}_2 $, i.e.~the center is projected out.

\begin{figure}
\begin{minipage}{7cm}
\centerline{\epsfxsize=7 cm\epsffile{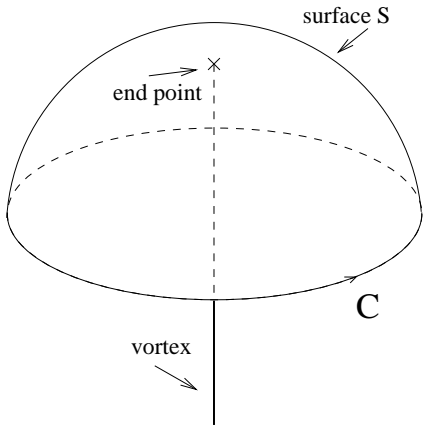}}
\caption{\label{loop}\textsl{Illustration of the closedness of center
vortices.}}
\end{minipage}
\end{figure}

From the considerations above we can also conclude that center vortices 
cannot have a boundary - they 
have to form closed lines or surfaces. 
Let us assume there is an end-point of a center vortex. We consider a
closed loop $\CC$ around the center vortex near the end-point. Then
there exists a surface $\CS$ bounded by $\CC$ such that the center
vortex does not intersect $\CS$, see fig. \ref{loop}. 
Consider now the double-valued
covering of $\CS$ given by the set of matrices 
$ \{V(x)g(x), -V(x)g(x)\} \, , \, x\in \CS$ fulfilling 
(\ref{cond1},\ref{cond2}). 
Because $\CS$ is contractible (i.e.~$\pi_1(\CS)=\{ 1 \}$) 
the covering of $\CS$ is topologically given by 
$\CS \times {\mathbb Z}_2$. 
Hence, if we lift  the loop $\CC$ into the bundle $P_C$ the lifted loop
is well defined and has no jumping points. But this contradicts our
assumption that there is a center vortex with an end-point.  

Especially center vortices cannot end at 
monopoles, i.e.~if there is one center vortex going in, there must be a 
second center vortex going in the monopole. On the other hand 
Dirac sheets can be open, i.e.~bounded by monopole loops.

\mysection{Specific Abelian and center gauges}
\label{examples}

Above we have considered Abelian and center gauge fixing from a
general point of view using two ``Higgs'' fields, i.e.~colour fields
living in the algebra of the gauge group and transforming covariantly
(homogeneously) under gauge transformations. Now we will consider
specific choices of these Higgs fields. We will start by considering the
continuum version of the Laplacian gauge. 

\subsection{Laplacian Abelian and center gauge}

In the Laplacian gauge the Higgs fields $\psi_1, \psi_2$ are identified
with the eigenfunctions of the two lowest eigenvalues of the covariant
Laplace operator \cite{vdSijs-96,Alexandrou-99-1,deForcrand-00}
\begin{equation}
\label{laplace-op}
- \hat \D_\mu (A) \hat \D_\mu (A) \psi_i = \lambda_i \psi_i \, ,
\end{equation}
where
\begin{equation}
\hat \D_\mu (A) \psi = [ \D_\mu (A) , \psi ] \, .
\end{equation}
As the Laplace operator is positive semidefinite all eigenvalues are
non-negative, i.e.~$\lambda_i \geq 0$. The eigenvectors $\psi_i(x)$
transform covariantly under gauge transformations and, in principle, we
could use any two eigenvectors $\psi_i , \psi_k $ as the Higgs fields
for Abelian or center gauge fixing. However, the spirit of the Abelian
and center gauge fixing is to extract the infrared degrees of freedom as
gauge fixing defects. For this purpose one should use the lowest lying
eigenvectors as Higgs fields since they carry the infrared content of a
gauge field configuration $A_\mu (x)$. 

In the following we study the effect of Abelian and center Laplacian
gauge fixing for specific field configurations.

\subsubsection{Laplacian Abelian gauge}
  
To demonstrate the emergence of magnetic monopoles in the Laplacian
Abelian gauge \cite{vdSijs-96} 
fixing let us consider the following gauge potential
\begin{equation}
\label{example1}
A_i = 0 , i=1,2,3 \, , \,  A_0 = a(r) \frac{x_k}{r} T_k \, ,  \, 
r^2 = x_1^2 + x_2^2 + x_3^2 \, , \, a(r) = \sin ( \pi r/R  ) \, , \, 
T_k = \sigma_k / (2 i)    
\end{equation}
on the space time manifold $D_3 \times S^1 $ (three dimensional ball with 
radius $R$ 
times circle with circumference $\beta$). 
As boundary conditions we demand periodicity in time and $ \psi_1 = 0$ 
on the boundary of $ D_3 $, implying there is a magnetic monopole on the
boundary of $D_3$. The Laplace operator acting on $ \psi_1 =
\psi_1^k \sigma_k $ reads
\begin{equation}
\label{laplace1}
- \hat \D^2 \psi_1 = 
- \left( \partial_i \partial_i \delta_{k l} + \partial_0^2 \delta_{k l}
+ 2 a \frac{x_j}{r} \varepsilon_{jkl} \partial_0  
- a^2 \delta_{kl} + a^2 \frac{x_k x_l}{r^2} \right) \psi_1^k \sigma_l \, .
\end{equation}
The ground state wave function is time independent and of hedgehog type, i.e.
\begin{equation}
\label{hedgehog1}
\psi_1^k = \frac{x^k}{r} b(r) \, .
\end{equation}
The function $b(r)$ has to fulfill the 
differential equation
\begin{equation}
r^2 b'' + 2 r b' + ( \lambda_1 r^2 - 2 ) b = 0 \, ,
\end{equation}
where $ \lambda_1 $ is the non-negative eigenvalue of the ground state
of the covariant Laplacian (\ref{laplace1}). 
The solution to this equation is
$$
b(r) = C \frac{1}{\sqrt{\lambda_1} r} \left( 
\frac{\sin (\sqrt{\lambda_1} r)}{\sqrt{\lambda_1} r} - 
\cos (\sqrt{\lambda_1} r ) \right) \, .
$$ 
The minimal eigenvalue $\lambda_1 $ is defined by the boundary condition
$ b(R) = 0 $, i.e.~$ \lambda_1 \approx (4.4934 / R )^2  $.

Abelian gauge fixing (\ref{cond1}) implies here to rotate the color
vector defined by the ground state (\ref{hedgehog1}) of the covariant 
Laplace operator
%\footnote{The field $\psi_1$ also fulfills the 
%Polyakov gauge condition and the normalized field fulfills the 
%maximally Abelian gauge condition as well.} 
into the $3$-direction and 
results in a static monopole line at $ r=0 $ (cf. equations 
(\ref{hedgehog},\ref{monopole-gauge-pot})). 
With the gauge transformation (\ref{monopole-gauge-trf}) we get 
a Dirac string on the negative $z$-axis. 
It connects the location of 
the magnetic monopole at $r=0$ with the one on the boundary of $D_3$.

\subsubsection{Laplacian center gauge}

As an example we will consider a thick vortex along the $z$-axis in 
three dimensional 
space $D_2 \times S_1$ (two dimensional disc with radius $D$ times circle with 
radius $Z$):
\begin{equation}
\label{thick-vortex}
A = f(\rho) \d \varphi T_3 
\, , \, f(\rho=0) = 0 , f(\rho=D) = 1 \, ,
\end{equation}
where we used cylinder coordinates. 
The gauge potential is invariant under rotation around the $z$-axis and under 
translation along the $z$-axis. 
For the Laplacian center gauge fixing we 
choose the two ``Higgs'' fields $\psi_{1,2}$ to be given by the ground
state and first excited state, respectively, of the covariant Laplacian.
Thereby we impose the boundary conditions that the Higgs fields 
vanish on the surface of the cylinder(for $\rho=D$) and are periodic 
in the $z$-direction. The covariant Laplace operator on 
$ \psi_{1,2} = \psi_{1,2}^k \sigma_k $ reads
\begin{eqnarray}
\hat \D_i \hat \D_i \psi_{1,2} &=& 
(\frac{1}{\rho} \partial_\rho \rho \partial_\rho 
+ \frac{1}{\rho^2} \partial_\varphi^2 + \partial_z^2) \psi_{1,2}^k \sigma_k +
\frac{f(\rho)^2}{\rho^2} \psi_{1,2}^k [T_3,[T_3,\sigma_k]] + \\
&& 2 \frac{f(\rho)}{\rho^2} \partial_\varphi \psi_{1,2}^k [T_3,\sigma_k]  .
\end{eqnarray}
Setting
\begin{equation}
\psi_{1,2}^c = \psi_{1,2}^1 + i \psi_{1,2}^2
\end{equation}
we get the following differential equations:
\begin{eqnarray}
(\frac{1}{\rho} \partial_\rho \rho \partial_\rho 
+ \frac{1}{\rho^2} \partial_\varphi^2 + \partial_z^2) \psi_{1,2}^3 
&=& - \lambda \psi_{1,2}^3 \\
(\frac{1}{\rho} \partial_\rho \rho \partial_\rho 
+ \frac{1}{\rho^2} \partial_\varphi^2 + \partial_z^2) \psi_{1,2}^c 
- \frac{f(\rho)^2}{\rho^2} \psi_{1,2}^c 
+ 2 \frac{f(\rho)}{\rho^2} i \partial_\varphi \psi_{1,2}^c 
&=& - \lambda \psi_{1,2}^c . 
\end{eqnarray}
If we make a separation ansatz for $ \psi_{1,2}^3 $ we get 
the Bessel differential equation for its $\rho$-dependence.
The ground state is then given by
$\psi_1^c = 0$ and 
$ \psi_1^3 = J_0 ( \sqrt{\lambda} \rho ) $, 
where $ J_0 $ is the zeroth Bessel function.
The minimal eigenvalue $\lambda$ is defined by the boundary condition
$J_0 ( \sqrt{\lambda} D ) = 0 $,  i.e.~$\lambda_1 \approx ( 2.4048 / D )^2 $.
The first excited state is given by
$ \psi_2^3 = 0 $ and
$ \phi_2^c = g(\rho) \exp ( - i \varphi ) $, where 
$ g $ has to fulfill the differential equation
\begin{equation}
\label{first-excited}
\rho^2 g'' + \rho g' + ( \lambda \rho^2 - ( 1 - f )^2 ) g = 0 .
\end{equation}
As an example let us take 
\begin{equation}
\label{example2}
f(\rho) = 1 - \exp ( -\rho / (1-\rho/D) ) , D=10 \, .
\end{equation}

\begin{figure}
%\begin{center}
\begin{minipage}{7cm}
\centerline{\epsfxsize=6.5 cm\epsffile{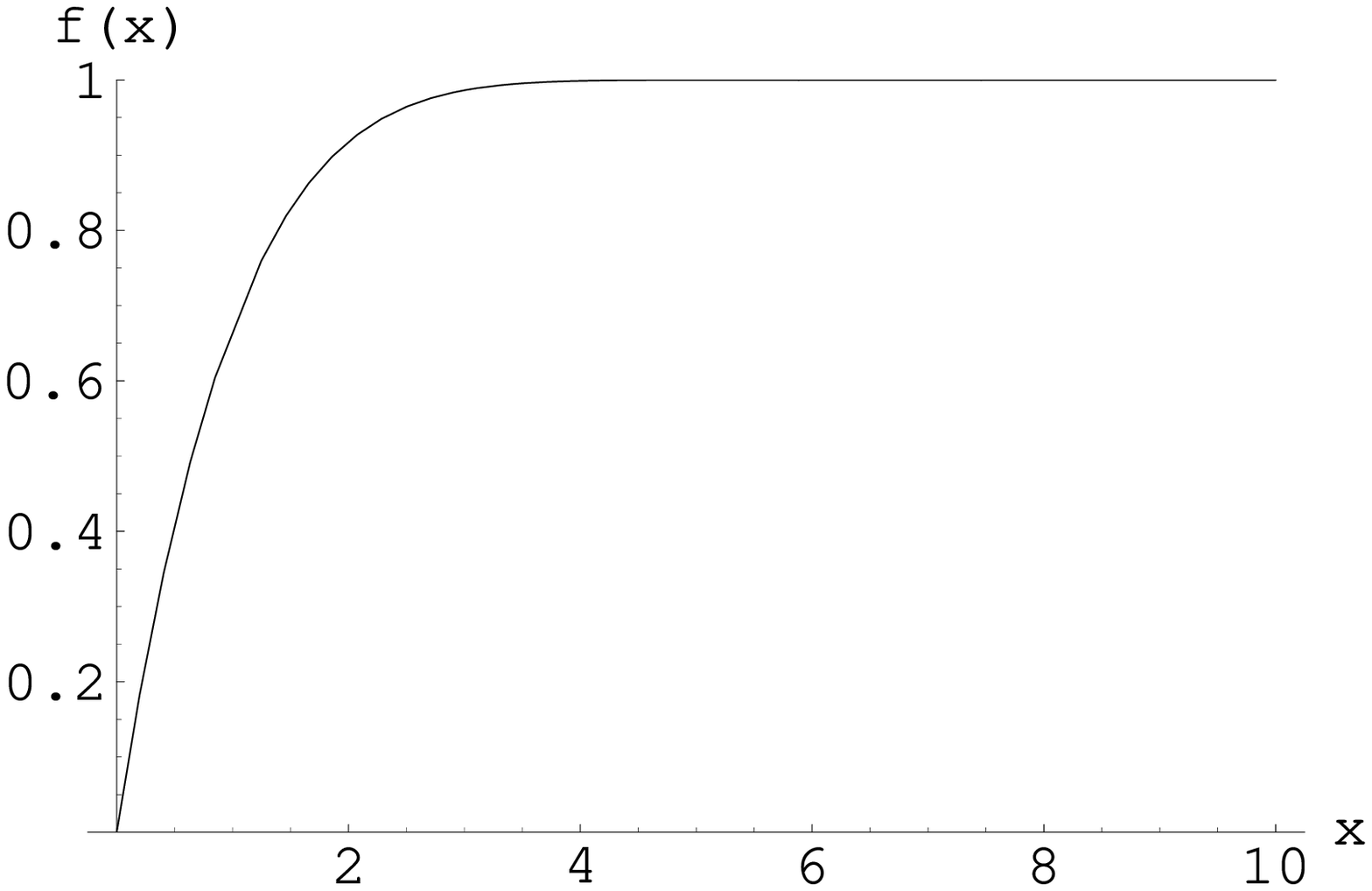}}%function1.eps}}
\caption{\label{function1}\textsl{The vortex profile function f, cp.
(\ref{example2}).}}
\end{minipage}
\hspace{1cm}
\begin{minipage}{7cm}
\centerline{\epsfxsize=6.5 cm\epsffile{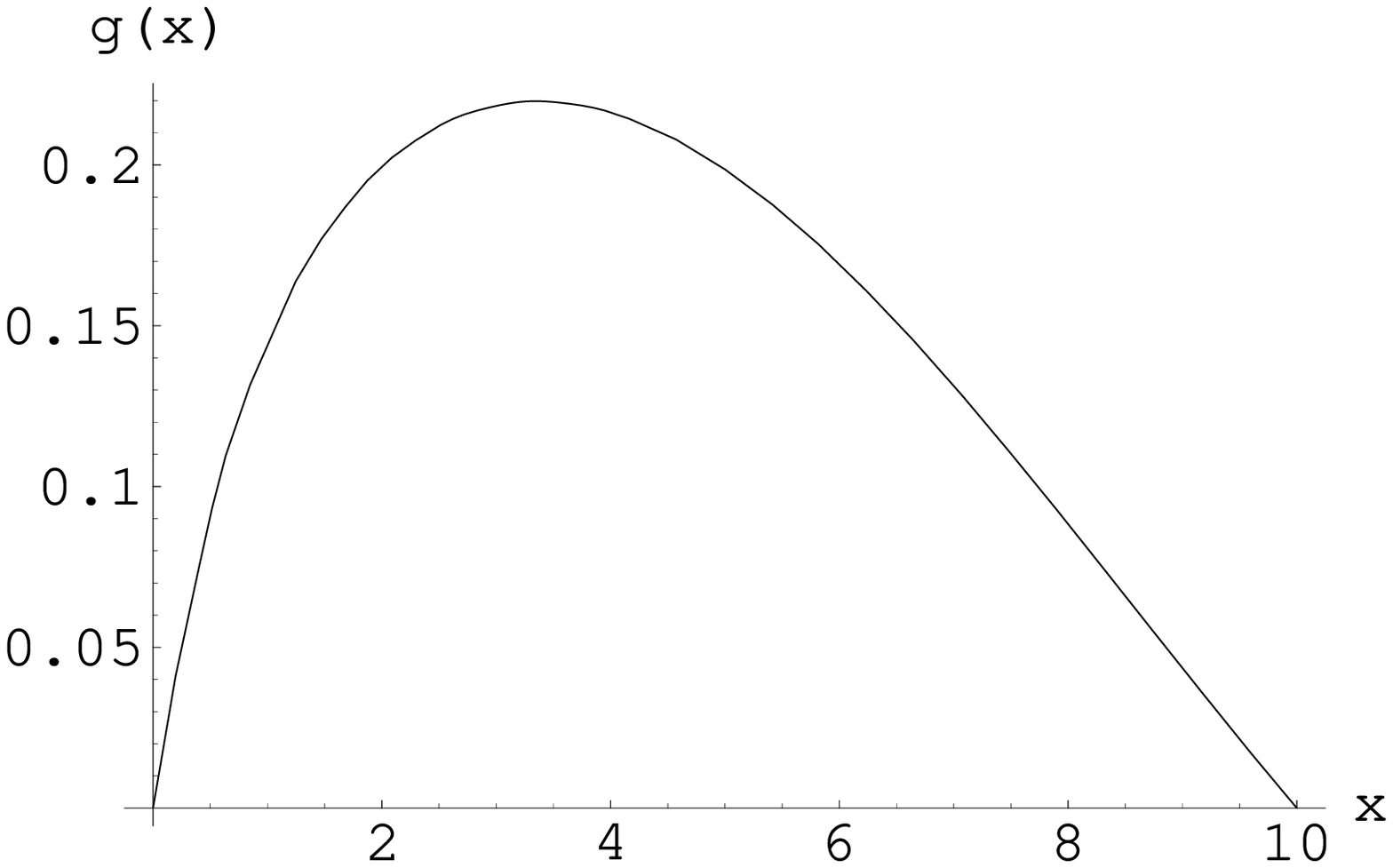}}%function2.eps}}
\caption{\label{function2}\textsl{The function g for the first excited
state, cp. (\ref{first-excited}).}}
\end{minipage}
%\end{center}
\end{figure}

The eigenvalue of the first excited state is
$\lambda_2 \approx  0.0835062 $ and 
its wave function has the form $ \psi_2 (\rho , \varphi ,z )  = 
g(\rho) ( \sigma_1 \cos \varphi - \sigma_2 \sin \varphi) $.  
After center gauge fixing this configuration leads to a center vortex 
on the $z$-axis, i.e.~subsequent Abelian projection (replacing 
the full gauge potential by its Abelian part) results in a ``thin'' 
vortex
\begin{equation}
A_{{\mathrm mag}} = \d \varphi T_3 \, .
\end{equation}
Thus in the continuum Abelian projection after Laplacian center gauge
fixing converts a thick center vortex into a thin one. This is analogous 
to what happens in center projection after maximal center gauge fixing 
on the lattice \cite{Engelhardt-00-1}.

\subsection{Abelian and center gauges from Wilson lines}

Here we consider space-time to be given by a $4$-torus and 
define fields $\phi_1$ and $\phi_2$ as path ordered 
exponentials of the gauge potential around different circumferences 
of the torus. But first we have to remind some facts about gauge fields
on the torus. 

We consider the torus as ${\mathbb R}^4$ modulo 
a discrete lattice spanned by 4 orthogonal vectors 
$b_\mu$, $\mu = 1, \ldots , 4$, having lengths $L_\mu= |b_\mu|$.
A gauge potential on the torus ${\mathbb T}^4$ is then given by a gauge 
potential on  ${\mathbb R}^4$ fulfilling periodicity properties:
\begin{equation}
\label{periodicity1}
A(x + b_\mu) = 
U_\mu^{-1} (x) A(x) U_\mu (x) +  U_\mu^{-1} (x) \d U_\mu (x) \, ,
\end{equation}
where the $U_\mu$ are called transition functions and they fulfill 
the Co-cycle condition:
\begin{equation}
\label{cocycle}
U_\mu(x) U_\nu(x+b_\mu) = U_\nu(x) U_\mu(x+b_\nu) \, .
\end{equation}
Equ. (\ref{periodicity1}) means that the gauge potential 
at the point shifted by $b_\mu$ is a gauge transformation of the original 
gauge potential. Therefore all (gauge independent) observables are periodic 
on $ {\mathbb R}^4$ - so they are well defined on ${\mathbb T}^4$.
Under a gauge transformation, $ V(x)$, the pair $(A,U)$ is mapped to
\begin{eqnarray}
\label{gauge-transf-A}
A^V (x) &=& V^{-1} (x)A(x) V(x) + V^{-1}(x) \d V(x) \, ,\\
\label{gauge-transf-U}
U_\mu^V (x) &=& V^{-1} (x) U_\mu (x) V(x+b_\mu) \, .
\end{eqnarray}
Now we define the two fields
\begin{eqnarray}
\nonumber
\phi_2 (x_0,x_1,x_2,x_3) &:=& 
\CP \exp \left( - \int_{(x_0,0,x_2,x_3)}^{(x_0,x_1,x_2,x_3)} A \right)
U_1 (x_0,0,x_2,x_3) 
\CP \exp \left( - \int_{(x_0,x_1,x_2,x_3)}^{(x_0,L_1,x_2,x_3)} A \right)
\, ,
\\
\phi_1 (x_0,\vec x) &:=& 
\CP \exp \left( - \int_{(0,\vec x)}^{(x_0,\vec x)} A \right)
U_0 (0,\vec x) 
\CP \exp \left( - \int_{(x_0,\vec x)}^{(L_0,\vec x)} A \right) \, , 
\end{eqnarray}
where $\CP \exp \int_a^b A$ denotes path ordered integration along a straight 
line from $a$ to $b$. Noticing 
\begin{equation}
\CP \exp \int_a^b A^V = V^{-1}(b) \CP \exp \left(\int_a^b A \right) V(a)
\end{equation}
and (\ref{gauge-transf-U}) we observe that $ \phi_{1,2}$ transform in the 
adjoint 
representation, i.e.~${\phi_i}^V (x) = V^{-1}(x) \phi_i (x) V(x)$ and 
they fulfill the correct periodicity properties, i.e.
$ \phi_i (x+b_\mu) = U_\mu^{-1}  (x) \phi_i(x) U_\mu(x)$.
%Abelian gauge fixing with $\phi_1 $ results in the well known Polyakov gauge
%on the torus, see \cite{Jahn-98,Reinhardt-97-2,Ford-98-1}. 

We could straightforwardly identify $\log \phi_1 = \psi_1$ and $\log
\phi_2 = \psi_2$ with the Higgs fields in the adjoint representation
introduced above to define the Abelian and center gauge fixing. Instead
we can work with the group valued fields $\phi_1$ and $\phi_2$
themselves. The Abelian gauge defined by rotating the vector $\psi_1^a$
into the $3$-direction corresponds to the diagonalization of $\phi_1$,
which defines the well known Polyakov gauge. Monopoles emerge now at
those isolated points $\vec x_i$ in space where the diagonalization of
$\phi_1$ is ill-defined, i.e.~where $\phi_1(\vec x_i)$ takes values in
the center of the gauge group ($\phi_1(\vec x_i) = \pm 1$ for $SU(2)$).

The corresponding center gauge fixing is still defined by rotating
$\psi_2^a = (\log \phi_2)^a$ into the $1$-$3$-plane, or equivalently to
rotate away the $\phi_2^{a=2}$ component. 
The desired gauge transformation $V g$ is defined by
\begin{eqnarray}
\phi_1^{Vg} (x) = g^{-1}(x) V^{-1} (x) \phi_1 (x) V (x)g (x)  
&=& \cos \alpha + i \sin \alpha \sigma_3 , \, 0 \leq \alpha \leq \pi , 
\\
\nonumber
\phi_2^{Vg} (x) = g^{-1} (x) V^{-1} (x) \phi_2(x) V(x) g(x) 
&=& \cos \beta + i \sin \beta \cos \gamma \sigma_3 
+ i \sin \beta \sin \gamma \sigma_1 , \, 0 \leq \beta \leq \pi .
\end{eqnarray}
Center vortices are identified when the color vectors
$(\log \phi_1)^a$ and $(\log \phi_2)^a$  are linearly dependent
(parallel or antiparallel) which translates to the condition that
$\phi_1$ and $\phi_2$ commute.

The above introduced center gauge can be considered as an extention of
the (Abelian) Polyakov gauge in the spirit of Palumbo gauges
\cite{Palumbo}.

\mysection{Merons and instantons in Laplacian center gauge}
\label{merons}

Of specific interest are instanton configurations since they dominate
the Yang-Mills functional integral in the semiclassical regime. Moreover
these objects carry non-trivial topological charge and are considered to
be relevant for the spontaneous breaking of chiral symmetry and for the
emergence of the topological susceptibility which by the
Witten-Veneziano formula \cite{Witten,Veneziano} provides the anomalous 
mass of
the $\eta '$. The instantons can, however, not account for confinement.
Early investigations have introduced merons to explain confinement,
which roughly speaking, can be interpreted as half of a zero-size
instanton (see below). In view of the recent lattice results supporting
the vortex picture of confinement 
\cite{DelDebbio-97,DelDebbio-98,Forcrand-99-1} 
merons should have some relation to center vortices if they, by any 
means, give rise to confinement. Furthermore meron pairs behave like
instantons concerning the chiral properties (see ref.~\cite{Steele} and
references therein).

\subsection{Merons as vortex intersection points}

In the following we will provide evidence that the merons can be
interpreted as vortex intersection points. We will then bring these
merons in the Laplacian center gauge and in fact detect a center vortex.

Merons are topologically non-trivial field configurations defined by 
\begin{equation}
\label{meron-gauge-pot}
A_{{\mathrm M}} %= {A_{{\mathrm M}}\,}^a_\mu (x) \d x_\mu T_a 
= \eta^a_{\mu \nu} \frac{x_\nu}{r^2} \d x_\mu T_a \, , \quad 
r^2 = x_1^2 + x_2^2 + x_3^2 + x_0^2 \, ,
\end{equation}
which possess Pontryagin index $\nu = \frac{1}{2}$. They can be considered as
half an instanton of vanishing radius. This becomes clear, if one compares the
gauge potential of the meron (\ref{meron-gauge-pot}) 
with the gauge potential of an instanton
\begin{equation}
\label{inst-gauge-pot}
A_{{\mathrm I}} = %{A_{{\mathrm I}}\,}^a_\mu (x) \d x_\mu T_a
= 2 \eta^a_{\mu \nu} \frac{x_\nu}{r^2 + \rho^2} \d_\mu T_a \, .
\end{equation}
Furthermore, the vanishing of the radius of the meron implies, that the
topological density of the meron is localized at a single point
\begin{equation}
Q (x) = \frac{1}{2} \delta^4 (x) \, .
\end{equation}
Obviously the meron has the same topological properties as an
transversal intersection point of two ${\mathbb Z}_2$ center vortex sheets
\cite{Engelhardt-00-1}. 
In the following we will show, that the
meron, in fact, shows all the features of an intersection point of vortex
sheets. We prove this by considering the Wilson loops around the 
center of the
meron. In fact, we will show, that for each color component the meron looks
near its center like a pair of intersecting ${\mathbb Z}_2$ center planes. 
For this we show, that the
Wilson loops in the corresponding planes yield center elements. To be more
precise we will show, that for a color component $b$ the Wilson loops around
the center of the meron in the plane $(i,j)$ and in the plane $(b,0)$ yield 
center elements, where the triplet of indices $(b,i,j)$ is defined by 
$| \epsilon_{b i j} | = 1$.

Consider a spherical Wilson loop $\CC$ in the spatial plane $(i,j)$. 
We can use polar coordinates in this plane
\begin{equation}
x_i = \rho \cos \varphi \, , \quad 
x_j = \rho \sin \varphi \, , \quad 
x_b = 0 = x_0 \, , \quad 
| \epsilon_{b i j} | = 1 \, .
\end{equation}
From the properties $\eta^a_{k l} = \epsilon_{a k l} $ and 
$ \eta^a_{0 k} = \delta^a_k $ of the t'~Hooft symbol $\eta^a_{\mu \nu}$
it follows that only the $b$-component in color space of 
$A_{{\mathrm M}}$ contributes to the Wilson loop, i.e.~the calculation
of the path ordered integral simplifies to ordinary integration of 
$A_{{\mathrm M}}$ along the path $\CC$.
\begin{eqnarray}
\oint\limits_\CC A_{{\mathrm M}} 
& = & 
\oint\limits_\CC \d x_\mu \eta^a_{\mu \nu} x_\nu T_a 
\frac{1}{x^2} 
\nonumber
\\
& = & 
\oint\limits_\CC \d \varphi \dot{x}_\mu (\varphi) 
\eta^a_{\mu \nu} x_\nu
(\varphi) \frac{1}{\rho^2} T_a \\ 
%\end{eqnarray}
%The integrand is different from zero only if the indices $(\mu, \nu)$ take the
%values $(i, j)$ or the permutation of it. Due to the properties 
%$\eta^a_{i j} = \epsilon_{a i j}$ of the 't~Hooft symbol 
%this fixes the color component.
%Hence, we obtain
%\begin{equation}
& = & \int^{2 \pi}_0 \d \varphi \epsilon_{a k l}
 \dot{x}_k x_l \frac{1}{\rho^2} T_a \, ,
\end{eqnarray}
where the indices $k, l$ run over the values $i, j$ and the integrand is
different from zero only for the color
component $b$. Straightforward
evaluations yield that the integrand is independent of the angle
$\varphi$, so that we eventually obtain
\begin{equation}
\oint\limits_\CC A_{{\mathrm M}} 
= - 2 \pi  \epsilon_{a i j} T_a \, .
\end{equation}
Hence, we find for the Wilson loop
\begin{eqnarray}
W (\CC) 
& = & 
\CP \exp \left({- \oint\limits_\CC A_{{\mathrm M}}}\right) 
= e^{+ \frac{i}{2} \sigma_a 2 \pi \epsilon_{a i j} } = -1
%\nonumber
%\\
%& = & 
%- 1 
%\quad \mbox{for} \quad \left|
%\epsilon_{b i j} \right| = 1 \, .
\end{eqnarray}

The lesson from this calculation is that for the meron 
only the color component 
$b$ defined by $\left| \epsilon_{b i j} \right| = 1$ contributes 
to the Wilson loop in the $(i, j)$ plane. Furthermore this Wilson loop
equals a center element, which can be interpreted by saying that the
$b$-component of the meron looks like a center vortex piercing the
$(i,j)$-plane with $|\epsilon_{b i j}| = 1$.

Let us now also show, that a Wilson loop in the plane orthogonal to 
the $(i, j)$ plane defined by $\left| \epsilon_{b i j} \right| = 1$ 
also receives contribution only from the color
component $b$ and yields also a center element.  
Indeed, for  the Wilson loop in
the $(0, b)$ plane, which is orthogonal  to the $(i, j)$ plane 
due to the condition $\left| \epsilon_{b i j} \right| = 1$, we find, 
introducing in  this plane analogous
polar coordinates,
\begin{equation}
x_0 = \rho \cos \varphi \, , \quad 
x_b = \rho \sin \varphi \, , \quad 
x_i = 0 = x_j \, , \quad 
\left| \epsilon_{b i j} \right| = 1
\end{equation}
and using the property 
$ \eta^a_{0 k} = \delta_{a k} $ of the 't~Hooft symbol 
\begin{eqnarray}
\oint\limits_\CC A_{{\mathrm M}} 
& = & 
\oint\limits_\CC 
\left( \d x_0 x_k - \d x_k x_0 \right) \eta^a_{0 k}
\frac{1}{\rho^2} T_a
\nonumber
\\
& = & 
- \int_0^{2 \pi} \d \varphi \left( \dot{x}_0 (\varphi) x_k (\varphi) - 
\dot{x}_k
(\varphi) {x}_0 (\varphi) \right) \delta_{a k} \frac{1}{\rho^2} T_a
\nonumber
\\
& = & 
- \int^{2 \pi}_0 \d \varphi T_b = - 2 \pi T_b \, .
\end{eqnarray}
We observe, that for the Wilson loop in the $(0,b)$ plane only the color
component $b$ contributes. Thus, indeed the color component $b$ of the 
meron
field looks like the intersection point of two vortex sheets, one in the 
$(i, j)$ and the other in the $(0, b)$ plane, where these indices are related by
$\left| \epsilon_{b i j} \right| = 1$.

One also easily shows, that in the remaining planes like 
$(i, 0) \quad i \neq b$
or $(i, k)  \quad k \neq j$, $ \left| \epsilon_{b i j} \right| = 1$
the spherical Wilson loops of the $b$-component of $A_{{\mathrm M}}$ 
around the center of the meron becomes trivial
\begin{equation}
W (\CC) = 1 \, .
\end{equation}

The remaining two color components of the meron field also behave 
like intersection points of two
transversal vortex planes $(i,j)$ and $(b,0)$ 
 as defined by the condition $\left| \epsilon_{b i j} \right| = 1$.

Thus we have seen, that indeed near its center the meron looks like pairwise
intersecting orthogonal center vortex sheets.

Now we will analyze the vortex content of the meron in Laplace center gauge. 
For this purpose we consider the meron on a 4-dimensional sphere $S^4$ 
with radius $R$. On $S^4$ we use stereographic 
coordinates $x_\mu,\mu=1,\ldots,4$. 
In these coordinates the metric is
conformally flat and reads
\begin{equation}
\label{metric-S4-stereographic}
g_{\mu \nu} = \frac{4 R^4}{(r^2+R^2)^2} \delta_{\mu \nu} \, .
\end{equation}
The covariant Laplace operator on $S^4$ has the form
\begin{equation}
\label{laplace-S4}
\hat \D^2 = 
\frac{1}{\sqrt{g}} \hat \D_\mu \sqrt{g} g^{\mu \nu} \hat \D_\nu = 
\frac{(r^2+R^2)^4}{16 R^8} (\partial_\mu + A^a_\mu \hat T_a) 
\frac{4 R^4}{(r^2+R^2)^2} (\partial_\mu + A^b_\mu \hat T_b) \, ,
\end{equation}
where $g$ denotes the determinant of the metric, $ r^2 = x_\mu x_\mu $ 
and $ \hat T_a $ are the generators of the gauge group in the adjoint 
representation. Plugging (\ref{meron-gauge-pot}) 
into (\ref{laplace-S4}) results in
\begin{equation}
\label{laplace-meron}
\hat \D^2 = \frac{(r^2+R^2)^2}{4 R^4} \left( 
\partial_r^2 + \frac{3}{r} \partial_r - 
\frac{4}{r^2} {\vec L \,}^2 - 
\frac{4}{r^2+R^2} r \partial_r
-\frac{4}{r^2} \hat {\vec T} \cdot \vec L - 
\frac{1}{r^2} \hat {\vec T}^2 \right) \, , 
\end{equation}
where $L^a = - i/2 \eta^a_{\mu \nu} x^\mu \partial_\nu$ and 
$\hat T^a = \ad (\sigma^a /2)$. 
$\vec L$ is the set of generators of an $SU(2)$ subgroup
of the rotation group $SO(4)$ \cite{tHooft-76-2}. Introducing the
conserved angular momentum $ \vec J = \vec L + i \vec T $  the 
eigenfunctions of the covariant Laplace operator $\hat \D^2$
(\ref{laplace-S4}) can be written in the form 
\begin{equation}
\label{eigenfunction-form-S4}
\psi(x) = f(r) \vec Y_{(j,l)} ( \hat x ) \cdot \vec \sigma \, .
\end{equation}
Here $\hat x_\mu = x_\mu / r $ and 
$Y_{(j,l)}$ denote the spherical vector harmonics on $S^3$ defined by  
\begin{eqnarray}
\label{sph-vector-harm}
{\vec L \,}^2 \vec Y_{(j,l)} = l (l+1)  \vec Y_{(j,l)} \, , \, 
{\vec J \,}^2 \vec Y_{(j,l)} = j (j+1) \vec Y_{(j,l)} \, , \, 
\hat {\vec T \,}^2 (\vec Y_{(j,l)} \cdot \vec \sigma )
= t (t+1)  \vec Y_{(j,l)} \cdot \vec \sigma \, , \, 
\end{eqnarray}
with $t=1$.
Substituting $f(r) = (r^2 + R^2) \varphi(r) $ \cite{Bruckmann-00-2} 
simplifies the eigenvalue problem problem to 
\begin{equation}
\left( 
-\partial_r^2 - 
\frac{3}{r} \partial_r + 
2 \frac{(j(j+1)+l(l+1)-1)}{r^2} -
\frac{8 R^2}{(r^2+R^2)^2} 
\right) \varphi 
= 
\lambda \frac{4 R^4}{(r^2+R^2)^2} \varphi \, .
\end{equation}
To get the lowest eigenvalue we have to minimize $(j(j+1)+l(l+1)-1)$. 
This quantity becomes minimal for $j=l=1/2$ (since the singlet $j=l=0$ 
is excluded by selection rules for $t=1$, see (\ref{sph-vector-harm})). 
Therefore the ground state is 4-fold degenerate and the meron
configuration lies on the Gribov horizon for the Laplacian center gauge
fixing. The four eigenfunctions form the
fundamental representation of $SO(4)$. The corresponding spherical
harmonics are given by:
\begin{equation}
\label{spherical-harmonics-1/2-1/2}
\left\{ 
Y_{(1/2,1/2)}^k \, , \, k=1,\ldots,4
\right\} =  
\left\{
  \left( \begin{array}{c}
                 - \hat x_4 \\
                   \hat x_3 \\
                   \hat x_2 
          \end{array} 
  \right) ,
  \left( \begin{array}{c}
                 - \hat x_3 \\
                 - \hat x_4 \\
                 - \hat x_1   
          \end{array} 
  \right) ,
  \left( \begin{array}{c}
                 - \hat x_2 \\
                   \hat x_1 \\
                   \hat x_4
          \end{array} 
  \right) ,  
  \left( \begin{array}{c}
                   \hat x_1 \\
                   \hat x_2 \\
                   \hat x_3
          \end{array} 
  \right) 
\right\} \, .
\end{equation}
Taking for instance the 4th eigenvector as the ground state and the
3rd as the first excited state the monopole and vortex content is
as follows. We get a static monopole line at $x_1=x_2=x_3=0$ and the
vortex sheet is the $(3,4)$-plane. 
Another possible choice for the two
eigenstates of the covariant Laplacian would be
$\psi_1 = \varphi ( Y_{(1/2,1/2)}^1 + Y_{(1/2,1/2)}^2 )$ and 
$\psi_2 = \varphi ( Y_{(1/2,1/2)}^3 + Y_{(1/2,1/2)}^4 )$. In this case
we identify a magnetic monopole line in the $(1,2)$-plane given by 
$x_1 = x_2 \, , \, x_3 = x_4 = 0$ and three center vortex sheets 
given by $x_2 = x_1 \, , \, x_3 = -x_4 \, , \quad $
$ x_1 = x_4 \, , \, x_2 = - x_3$ and $x_1 = -x_4 \, , \, x_2 = x_3$, 
respectively. The three center vortex sheets 
intersect at the origin. The meron configuration is
$SO(4)$ symmetric and the eigenspace to the lowest eigenvalue of the
Laplace operator shares this symmetry. Therefore we can move the vortex
planes by arbitrary $SO(4)$ rotations (this corresponds to choosing
other linear combinations of the degenerate eigenstates 
(\ref{spherical-harmonics-1/2-1/2}) of the covariant Laplacian for the
gauge fixing).    

Let us emphasize that the Laplacian center gauge fixing of the meron
field detects either a single vortex sheet or three center vortex 
sheets, while the study of the Wilson
loop has revealed pairwise intersecting vortex sheets near the meron
center. Obviously highly symmetric configurations like the meron or
instanton fields are not faithfully reproduced by the center projection
implied by the vortex identification of the Laplacian gauge fixing.
This is because these configurations are lying on the Gribov horizon.

\subsection{Instantons in Laplacian center gauge}

Below we consider a simple instanton and an instanton-anti-instanton
pair in  the Laplacian center gauge in order to reveal its monopole and
center vortex content. In the Laplacian Abelian gauge (which represents
a partial gauge fixing of the Laplacian center gauge) a simple instanton
has been considered recently \cite{Bruckmann-00-2}. We will not stick to
the Abelian gauge but consider the full Laplacian center gauge. In
addition, we do not confine ourselves to a single instanton but consider
also an instanton-anti-instanton pair. Such a configuration has
previously been studied on the Lattice \cite{Alexandrou-99-2}. 
For a single
instanton due to its symmetry the lowest lying eigenvectors of the
Laplacian can be found analytically when choosing $S^4$ as space-time
manifold \cite{Bruckmann-00-2}.

\subsubsection{The single instanton in Laplacian center gauge}

As in the above discussed meron configuration we use stereographic
coordinates $x_\mu$ on $S^4$ and the metric
(\ref{metric-S4-stereographic}). With the instanton gauge potential 
(\ref{inst-gauge-pot}):
$$
A_{{\mathrm I}} = 
2 \eta_{\mu \nu}^a \frac{x_\nu}{r^2 + \rho^2} \d x_\mu T_a
$$
the covariant Laplace operator reads  
\begin{equation}
\hat \D^2 = \frac{(r^2+R^2)^2}{4 R^4} \left( 
\partial_r^2 + \frac{3}{r} \partial_r - 
\frac{4}{r^2} {\vec L \,}^2 - 
\frac{8}{r^2 + \rho^2} \hat {\vec T} \cdot \vec L -  
\frac{4 r^2}{(r^2 + \rho^2)^2} \hat {\vec T \,}^2 -
\frac{4}{r^2+R^2} r \partial_r
\right) \, . 
\end{equation}
Again the eigenfunctions of $\hat \D^2$ have the form
(\ref{eigenfunction-form-S4}). Depending on the ratio $\rho/R$ 
between the scale $\rho$ of the instanton and the radius $R$ of
the 4-sphere the ground state is $3$-fold degenerate for $\rho \neq R$
and $10$-fold degenerate for
$\rho=R$. In the physical case $ R > \rho $ (including the 
infinite volume limit) the ground state is three-fold degenerate 
and has the form
\begin{equation}
\psi(x) = \frac{1}{R (R^2+r^2)} \vec Y_{(0,1)} \cdot \vec \sigma \, , 
\end{equation}  
i.e.~$j=0$ and $l=1$, see (\ref{sph-vector-harm}).
The triplet of functions $\vec Y_{(0,1)} $ is given by
\begin{equation}
\label{spherical-harmonics-0-1}
\left\{
  \left( \begin{array}{c}
                 \hat x_1^2 - \hat x_2^2 - \hat x_3^2 + \hat x_4^2\\
                 2 ( \hat x_1 \hat x_2 + \hat x_3 \hat x_4 ) \\
                 2 ( \hat x_1 \hat x_3 - \hat x_2 \hat x_4 ) 
          \end{array} 
  \right) ,
  \left( \begin{array}{c}
                 2 ( \hat x_1 \hat x_2 - \hat x_3 \hat x_4 ) \\
                 - \hat x_1^2 + \hat x_2^2 - \hat x_3^2 + \hat x_4^2 \\
                 2 ( \hat x_2 \hat x_3 + \hat x_1 \hat x_4 )  
          \end{array} 
  \right) ,
  \left( \begin{array}{c}
                 2 ( \hat x_1 \hat x_3 + \hat x_2 \hat x_4 ) \\
                 2 ( \hat x_2 \hat x_3 - \hat x_1 \hat x_4 ) \\
                 - \hat x_1^2 - \hat x_2^2 + \hat x_3^2 + \hat x_4^2 
          \end{array} 
  \right)   
\right\} \, .
\end{equation}
To get the monopole and vortex content of the configuration we have to choose 
one of the three eigenfunctions as the ground state and another as the
first excited state. But the only zeros of the eigenfunctions are at the 
origin. This means that the set of monopoles consists of the origin only, 
i.e.~we have no monopole loop or we can say that it is degenerated to a 
single point. 
To examine the vortex content of the configuration we have to look for
points where two of the three vectors in (\ref{spherical-harmonics-0-1})
are linearly dependent. But it is easy to see that the three vectors are
always perpendicular to each other. Therefore the set of vortices
consists also only of the origin.

For the special case $ R= \rho$ the ground state is $10$-fold
degenerate. In this case the set of ground states consists of 
two triplets ($j=0,l=1$ and $j=1,l=0$) and one quadruplet ($j=l=1/2$).
Choosing eigenfunctions from the quadruplet,
see (\ref{spherical-harmonics-1/2-1/2}),
as ground and first excited state we get the same result as in the meron
case, i.e.~a monopole line and one or three vortex sheets 
(see previous subsection).

\subsubsection{Instanton-anti-instanton pair in Laplacian center gauge}

We choose here space-time manifold as the direct product of a
three-dimensional disc $D^3$ with radius $D$ and an interval
$I=[-L_0,L_0]$. We consider a gauge potential describing approximately 
an instanton-anti-instanton pair: 
\begin{eqnarray}
\label{inst-gauge-potential}
 A_{{\mathrm {I A}}} &=& 2 \big( \eta_{\mu \nu}^a 
     ( x_\nu - z_\nu ) B_-
\nonumber
\\     
    && + \bar \eta_{\mu \nu}^a 
     ( x_\nu + z_\nu ) B_+
     \big) \d x_\mu T_a \, , \, T_a = \sigma_a / (2i) \, , 
\\
\label{profile-functions}
 B_\pm &=&  \frac{1}{|x \pm z|^2 + \rho^2} 
     \exp{(-(1.25 |x \pm z|/D)^{40})} \, .
\end{eqnarray}
The centers of the two instantons $ \pm z_\mu $ are located on the time
axis, i.e.~$ z_i = 0 \, , \, i = 1, 2, 3 $.
The Higgs fields (i.e.~the two lowest lying eigenfunctions) should 
vanish 
on the boundary of $D^3$ and be periodic in the time direction:
\begin{eqnarray}
\psi (x_0, \vec x ) &=& 0 \, \mbox{ for } \, |\vec x| = D \, , 
\\
\psi (-L_0, \vec x) &=& \psi (L_0, \vec x) \, .
\end{eqnarray}
The exponential factors in (\ref{profile-functions})  are introduced
to make the gauge potential nearly vanishing on the boundary of the 
space-time and to render the Laplace operator selfadjoint. 

\begin{figure}
\begin{minipage}{7cm}
\centerline{\epsfxsize=7 cm\epsffile{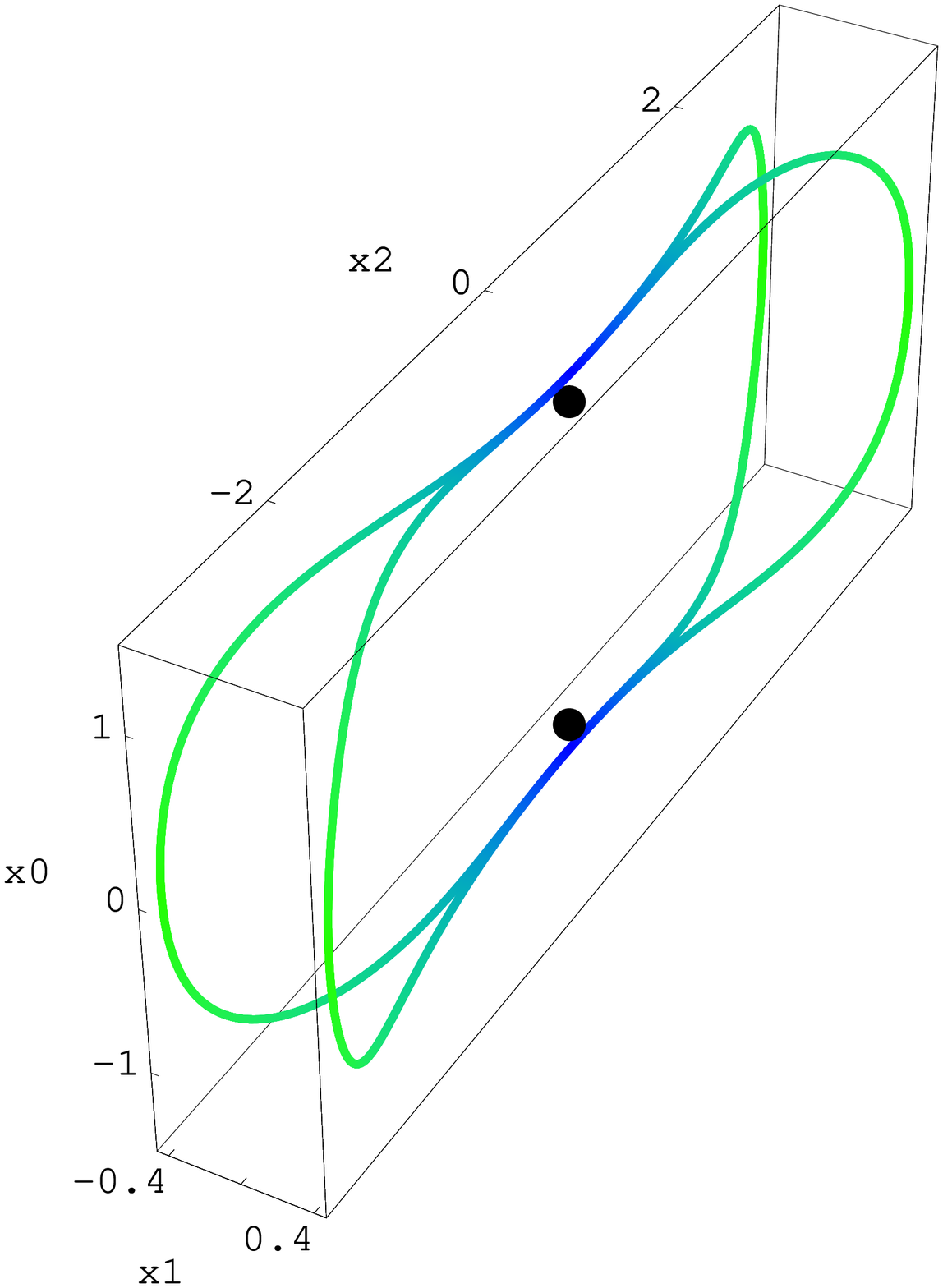}}
\caption{\label{monopoleloops}\textsl{Plot of the two magnetic monopole 
loops 
for the gauge potential (\ref{inst-gauge-potential}) projected onto the
$x_1-x_2-x_0$-space (dropping the $x_3$-component). 
Rotations with angle $\pi$ around the $x_1$- , $x_2$- and $x_3$-axis 
interchange the
different monopole branches. The thick dots show the positions of the
instantons.}}
\end{minipage}
\hspace{1cm}
\begin{minipage}{7cm}
\centerline{\epsfysize=8 cm\epsffile{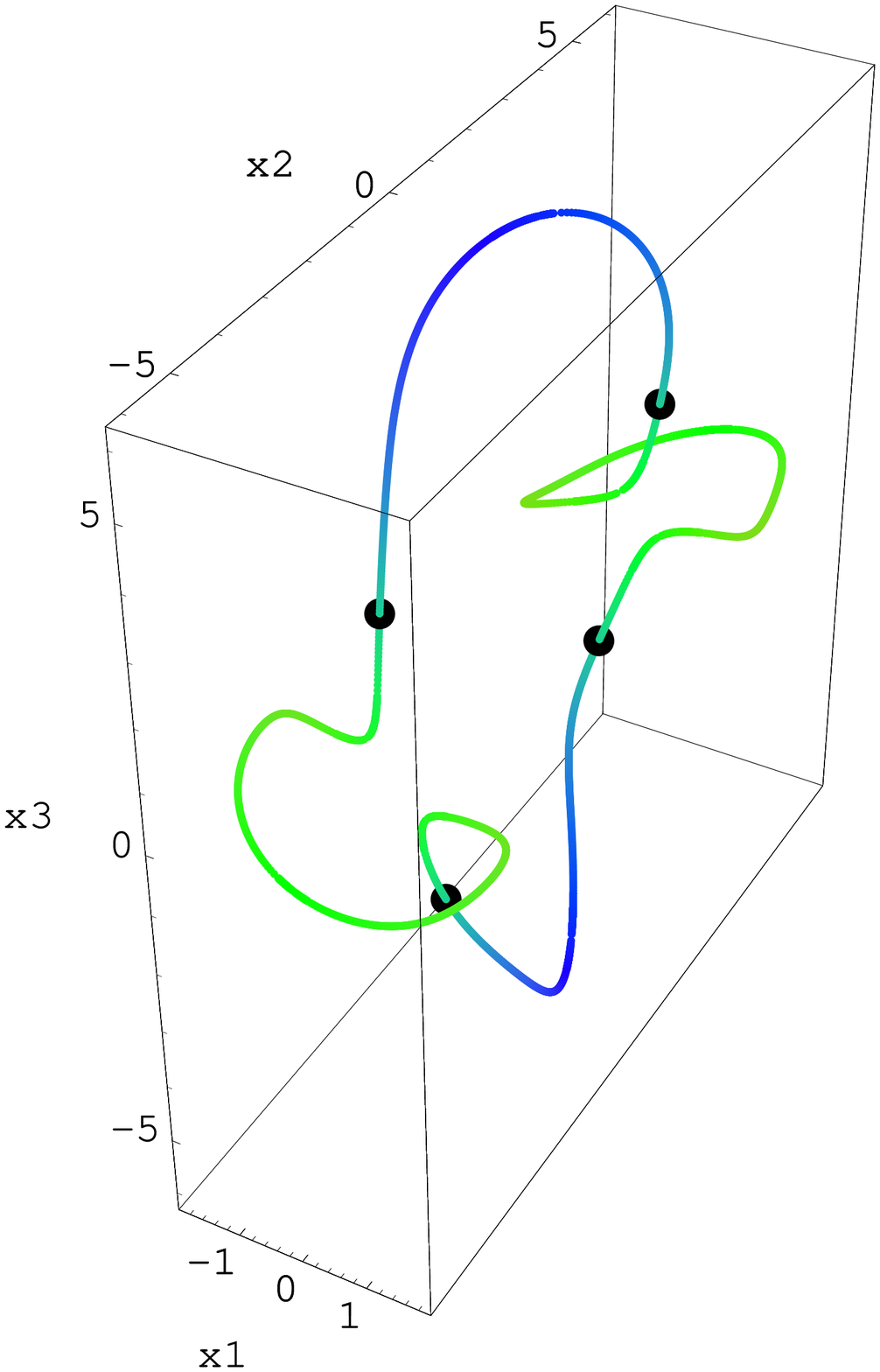}}
\caption{\label{vortex-picture}\textsl{Plot of the vortex in the
time-slice $x_0 = 0$. The thick dots on the vortex 
show the positions of the magnetic monopoles.}}
\end{minipage}
\end{figure}

For the considered instanton-anti-instanton configuration 
the Laplace operator reads
\begin{eqnarray}
\label{inst-laplace-op}
\hat \D_\mu \hat \D_\mu &=& \partial_{\mu} \partial_{\mu} 
    + 4 \left( B_+ - B_- \right) 
      \left( \vec x \cdot \hat {\vec T} \right) \partial_0 
\nonumber
\\    
    &&
    + 4 \left( (x_0 - z_0) B_- - (x_0 + z_0) B_+ \right) 
      \left( \hat {\vec T} \cdot \vec \partial \right) 
    + 4 \left( B_+ + B_- \right) 
      \hat {\vec T} \cdot \left( \vec x \times \vec \partial \right) 
\nonumber
\\
    &&
    + 4 \left( |x-z|^2 B_-^2 + |x+z|^2 B_+^2 + 
        2 ( r^2 - x_0^2 + z_0^2 ) B_- B_+ \right) 
      \left( \hat {\vec T} \cdot \hat {\vec T} \right)
\nonumber
\\       
    &&
    - 16 B_- B_+ \left( \vec x \cdot \hat {\vec T} \right) \cdot 
      \left( \vec x \cdot \hat {\vec T} \right) \, ,
\end{eqnarray}
where $r = |\vec x |$ and $ \hat {\vec T} = \ad (\vec T )$ is the color
spin in the adjoint representation.
The Laplace operator commutes with $ {\vec J \,}^2 $ and 
$ J_3 $, where $ \vec J = \vec L + i \vec T $ 
is the total angular momentum and 
$ \vec L = - i \vec x \times \vec \partial $ 
is the orbital angular momentum.
This means we can expand the eigenfunctions of $ \hat \D_\mu^2 $ in 
vector spherical harmonics $ \vec Y_{j l m} $  on $S^2$ \cite{Varshalovich}, 
with $ j(j+1) $, 
$ l(l+1) $ and $ m $ being the eigenvalues of $ {\vec J \,}^2 $,
$ {\vec L}^2 $ and $ J_3 $:
\begin{eqnarray}
\psi_{jm} = \sum_{l} T_{l} ( x_0 ) R_{l} ( r ) 
        \vec Y_{jlm} (\vartheta, \varphi ) \cdot \vec \sigma \, .
\end{eqnarray}
It turns out that the action of $\hat \D_\mu^2 $ on 
$\vec Y_{jlm} \cdot \vec \sigma $ 
does not depend on $m$. Therefore the eigenvalues of 
$\hat \D_\mu^2$ will be $(2 j +1 ) $-fold degenerate. 
The functions $ T_l(x_0) $ have been Fourier expanded in 
$ \sin $- and $ \cos $-functions of the time and 
$ R_l(r) $ in Bessel functions of $r$.
%The operator $ D_mu^2 $ does not mix between eigenfunctions with
%different values of $j$ and $m$, but it mixes functions with different
%values of $l$.
We solved the eigenvalue problem numerically by calculating the matrix
elements of $ \hat \D_\mu^2 $ and diagonalizing this matrix.
It turned out that the ground state has $ j=1 $ and thus it is 
threefold degenerate.
To get rid of the degeneracy we assume that we have an infinitesimal 
perturbation by $ \varepsilon J_3^2 $, such that the ground state has 
$m=0$.  We first consider a widely separated instanton and anti-instanton
configuration where the distance between the centers of instanton and
anti-instanton is large compared to the (anti-)instanton size $\rho$ 
(instanton and anti-instanton radii are chosen to be equal $\rho$).
For this case we have chosen the parameters as follows:
\begin{eqnarray}
D = L_0 = 10 \, , \, z_0 = 1 \, , \, \rho = 0.1 \, .
\end{eqnarray} 
From the zeros of the lowest eigenmode we identified two magnetic charge 1 
monopole loops crossing each other near the
instanton centers, see figure \ref{monopoleloops}. 
The set of the magnetic monopole loops is symmetric with respect 
to rotations with angle $\pi$ around the $x_1$-, $x_2$- and $x_3$-axis.

To identify the center vortices we have chosen 
$ \psi_{j=1,y} = i / \sqrt{2} ( \psi_{j=1,m=-1} + \psi_{j=1, m=1} ) $ 
as the first excited state. The resulting vortex connects at 
each time $ x_0 $ all four 
monopole branches, i.e~the vortex sheet is topologically equivalent to
$S^2$ and encloses the two instanton centers. 
In figure \ref{vortex-picture} we plotted the 
vortex  in the time-slice $ x_0 = 0 $.

Further we examined the dependence of the monopole and vortex content of the
configuration on the distance $ 2 z_0 $ between the instanton centers.
Reducing $z_0$ results in smaller monopole loops and at a critical value
($z_0 = 0.3513$) the monopole loops and the vortex sheet disappear.

\begin{figure}
\begin{minipage}{7cm}
\centerline{\epsfxsize=7 cm\epsffile{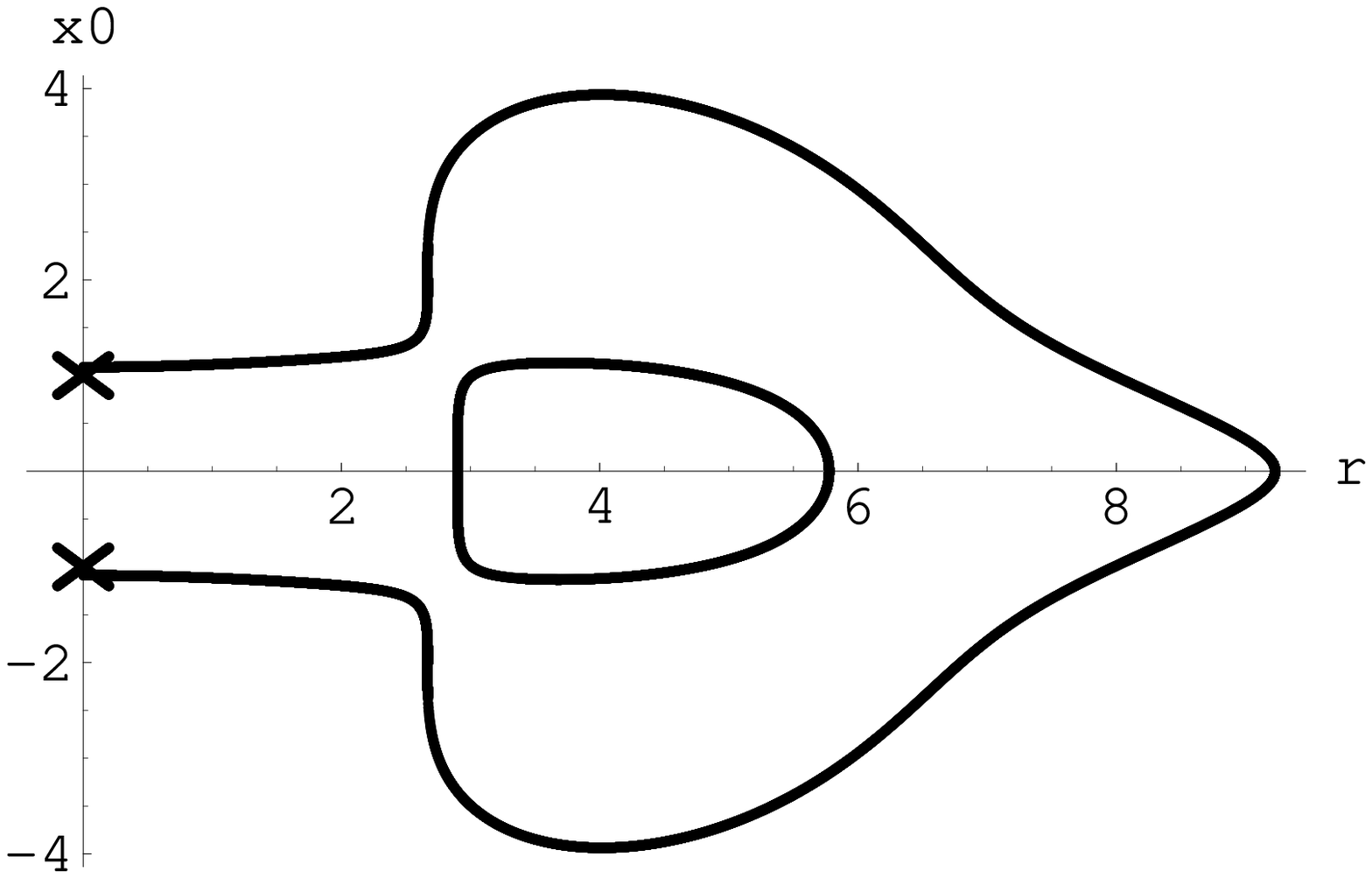}}
\caption{\label{monopole-2}\textsl{Plot of the $r-x_0$-projection 
($r=\sqrt{x_1^2+x_2^2+x_3^2}$) of 
one of the $4$ small magnetic monopole loops and one 
half of one of the $2$ large magnetic monopole loops for the gauge 
potential $2 A_{{\mathrm{I A}}}$. The crosses show the positions of the
instanton centers.}}
\end{minipage}
\hspace{1cm}
\begin{minipage}{7cm}
\centerline{\epsfysize=8 cm\epsffile{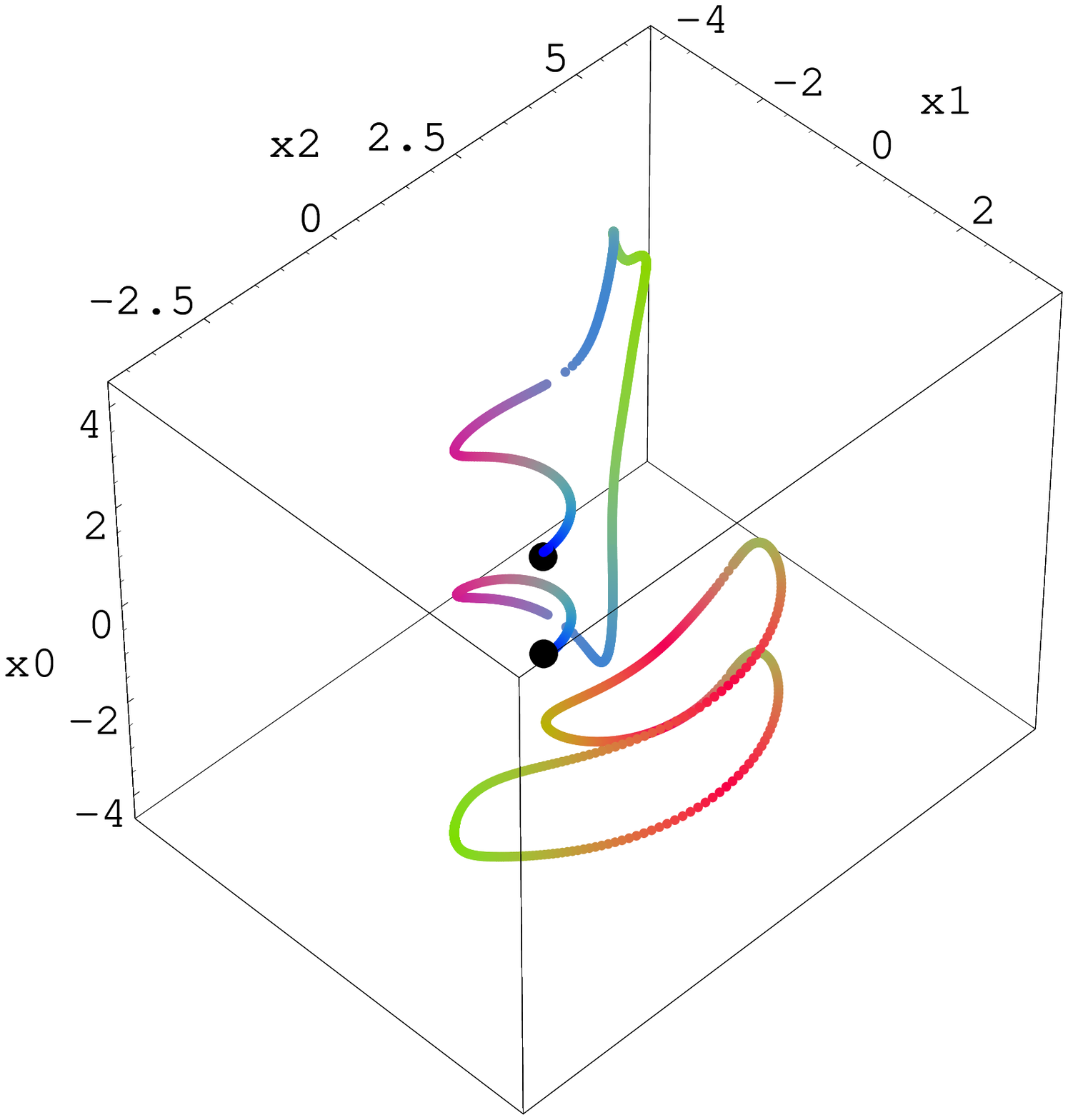}}
\caption{\label{monopole-2a}\textsl{$3$-dimensional plot of the
monopole loops for the doubled instanton-anti-instanton 
gauge potential projected onto the
$x_1-x_2-x_0$ space. Only one of the $4$ small magnetic 
monopole loops and one half of the $2$ large magnetic monopole 
loops are plotted. The
remaining part of the monopole loop is obtained by rotations with angle
$\pi$ around the $x_1$-, $x_2$- and $x_3$-axes, respectively.
The thick dots show the positions of the instanton centers.}}
\end{minipage}
\end{figure}

At the end we changed the gauge potential (\ref{inst-gauge-potential})
by a factor $2$. The result is a higher field
strength. Accordingly, after Laplacian Abelian gauge fixing, the number
of magnetic monopole loops increases. We identified $6$ magnetic
monopole loops 
--- two of them are larger and intersect each other on the $x_0$ axis 
(similar as in the case with gauge potential $A_{{\mathrm IA}}$,
cf.~(\ref{inst-gauge-potential})), while the other four monopole loops 
are smaller and separated from each other, cf.~figures 
\ref{monopole-2},\ref{monopole-2a}. The set of all magnetic monopole
loops is again symmetric with respect to rotations with angle $\pi$
around the $x_1$-, $x_2$- and $x_3$-axis.

\mysection{Concluding remarks}
\label{conclusion}

We have studied various field configurations relevant for the infrared
sector of QCD in the continuum analog of Laplacian (Abelian and) 
center gauges. While the gauge does not detect center vortices for
single instantons it identifies center vortices for merons and composite
instanton-anti-instanton configurations. The absence of center vortices
in single instantons is somewhat expected if center vortices are
responsible for confinement, which is, however, not explained by
instantons. Furthermore we have also shown that for highly symmetric
field configurations Laplacian center gauge does not necessarily provide
a very faithful method for detecting their vortex content, because these
configurations lie mostly on the Gribov horizon. A better
detector for center vortices is the Wilson loop. From the study of the
Wilson loop we have provided evidence that merons can be interpreted as
self-intersection points of center vortices.

\section*{Acknowledgements}

Diskussions, in particular on the numerics, with R.~Alkhofer, 
K.~Langfeld and A.~Sch\"afke we gratefully acknowledge.
Furthermore the authors thank Ph.~de~Forcrand for comments on the first
version of the manuscript and, in particular, for drawing our attention
to ref.~\cite{deForcrand-00}.

\end{document}